\documentclass[12pt]{iopart}

\usepackage{amssymb}

\usepackage{graphicx}
\usepackage{multirow}

\graphicspath{ {./images/} }
\usepackage{subcaption}
\usepackage{hyperref}
\hypersetup{
linkcolor=black,
citecolor=green
}

\begin{document}

\title[Modeling of charged-particle multiplicity and transverse-momentum ...]{Modeling of charged-particle multiplicity and transverse-momentum distributions in $pp$ collisions using a DNN}

\renewcommand{\thefootnote}{\fnsymbol{footnote}}

\author{{E. Shokr*\footnote[0]{*Corresponding author.}}}

\address{Physics Department, Faculty of Science, Mansoura University, Mansoura, Egypt}
\ead{eslam.shokr@cern.ch}

\author{A. De Roeck}
\address{CERN, Geneva, Switzerland}
\ead{Albert.de.Roeck@cern.ch}

\author{M. A. Mahmoud*}
\address{Center for High Energy Physics (CHEP-FU), Faculty of Science, Fayoum University, Egypt}
\ead{mohammed.attia@cern.ch}

\vspace{10pt}
\begin{indented}
\item[]\today
\end{indented}

\begin{abstract}
A machine learning technique is used to fit multiplicity distributions in high energy proton-proton collisions and applied to make predictions for collisions at higher energies. The method is tested with Monte Carlo event generator 
events.
Charged-particle multiplicity and transverse-momentum distributions within different pseudorapidity intervals in proton–proton collisions were simulated using the PYTHIA event generator for center of mass energies $\sqrt{s}$= 0.9, 2.36, 2.76, 5, 7, 8, 13 TeV for model training and validation and at 10, 20, 27, 50, 100 and 150 TeV for model predictions. Comparisons are made in order to  ensure the model reproduces the relation input variables and  output distributions for the charged 
particle multiplicity and transverse-momentum. The multiplicity and transverse-momentum distributions are described and predicted very well, not only in the case of the trained but also in the untrained energy values.
The study proposes a way to predict multiplicity distributions at a new energy by extrapolating the information inherent 
in the lower energy data. Using real data instead of Monte Carlo, as measured at the LHC, the technique has the potential to project the multiplicity 
distributions for different intervals at very 
high 
collision energies, e.g. 27 TeV or 100 TeV for the upgraded HE-LHC and FCC-hh respectively, using only  data collected at the LHC, i.e. at center of mass energies from 0.9 up to 13 TeV. 

\end{abstract}

%
%
%
%
%

\section{Introduction}
\label{}

Inclusive particle multiplicity distributions are among the most basic global characteristics of high 
energy proton-proton ($pp$) collisions\cite{Abel2010}, but have
been proven to be difficult to describe or predict by standard Monte
Carlo generator programs, such as 
PYTHIA\cite{PythiaPhysicManual} and HERWIG\cite{HerwigPhysicsAndManual}. 
The $pp$ charged-particle multiplicity 
has been studied  theoretically and experimentally at the Large Hadron Collider (LHC) in different experiments and for various colliding center of mass (CM) energies ($\sqrt{s}$) \cite{Abel2010, Aamodt2010_7, Khachatryan2011,Aad2011, Acharya2017, Adam2017, Aaij2017, our}. 
charged-particle multiplicity distributions generated in these collisions in restricted 
pseudo-rapidity intervals ($\Delta\eta$) , i.e. the probability $P(N_{ch},\sqrt{s},\Delta\eta)$ for the number of charged-particles in the final state ($N_{ch}$), depend on the number of interactions between quarks and gluons confined inside the colliding protons, and the underlying mechanisms of particle production 
\cite{grosse2010charged}.

At LHC energies, $pp$ interactions are dominated by soft QCD processes, i.e. interactions with small transverse-momentum ($p_{T})$ transfer. Such interactions  cannot be treated perturbatively but are modeled phenomenologically \cite{Khachatryan2010}. These processes are very useful for studying QCD in non-perturbative regimes, tuning event generators and constraining the dynamics in phenomenological models. As the collision energy increases, the contributions from hard scattering increase which can be treated perturbatively. A generic term for such events is minimum bias (MB) events, 
which is not a physics but an operational definition, depending on the minimum requirements imposed to select 
such an event (e.g. based on the amount of energy  or number of particles observable in 
the experiment). 

At the LHC, PYTHIA and HERWIG are the common generators to describe the $pp$ multiplicity distributions at
the various center of mass energies at which the collider has operated over the past years.
 Comparisons to data at the different CM energies show  that it is very challenging to describe the charged-particle multiplicity distributions with these models, despite the many tunable parameters available for the 
user. Moreover, we cannot be sure how well these parameters allow to  cover the underlying dynamics and its energy dependence 
of in particular these soft processes.
Sufficiently accurate descriptions of multiplicity distributions are however important at hadron colliders where
we can have, now and in the future, about 20 to perhaps a few hundred of such minimum bias events per bunch crossing  
overlapping with an event of interest.
These additional events add significantly to the occupancy in the detectors and affect systematical uncertainties of precision measurements. As soon as such future hadron colliders turn into operation the characteristics of MB events 
will be measured 
in a very early stage of the operation, 
but until then, studies on the capabilities of 
such new machine will have to rely on model predictions.

Therefore we present in this study an alternative approach where we make no prior assumption on any underlying
model or tuning of parameters, but use a machine learning technique to construct "the model". This is similar to the 
very successful parton density distribution (PDF) determinations technique used by the 
NNPDF collaboration~\cite{Ball:2008by}, where instead of imposing explicit functional forms for the distributions 
at a starting scale, a neural network is used to provide that information, in order to reduce the source of 
potential bias from the initial 
assumptions.

The $p_T$ spectrum of final state charged hadrons is also an important observable in describing particle production in $pp$ collisions \cite{Chatrchyan_2011}. As an example, the study of the $p_{T}$ spectrum in $pp$ collisions offers a reference for the measurements of the suppression of high-$p_T$ particles (Jet Quenching) in
a dense QCD medium produced in ion-ion collisions 
\cite{Suppression201130,Singh:2013fha}.
A  solid knowledge of the rates and
characteristics  of the particle production are mandatory to distinguish e.g. rare soft processes from the relatively huge backgrounds of hadronic interactions \cite{Farrar:2017eqq}, which is one of the greatest challenges in these pursuits, and
for extracting precision measurements from the 
data.

Since several years, particle physicists have continued to explore techniques to increase the analyzing power for
measurements by using algorithms implementing multiple variables simultaneously. 
These so-called 
multivariate analyses techniques  \cite{multi, wolter, lhcapp} 
have been shown to provide significant support for different challenges in data analysis but also have some important limitations, 
with increasing of the dimensionality of the problem.

The implementation of these advanced analysis techniques, such as Machine Learning (ML), the increasing computer power and tailored processors
for the problem, 
and the emergence of Deep Learning (DL) techniques around 2012 \cite{krizhevsky2012imagenet} allowed for tools to tackle previously limitations of handling higher-dimensional and simplifying more complex problems.
In high energy physics, machine learning algorithms and techniques have been embraced 
early on for analyzing and collecting the huge amount of data produced by colliders \cite{multi}; e.g. LHC is presently one of the largest data volume generators. The role of these new powerful techniques is clear, namely revolutionizing the handling and interpretation of these huge data volumes, and allowing to extract detailed 
 physics results with increased sensitivity. These techniques are
 now considered essential tools at the LHC and
 have found  important applications in data analyses, calibration, event triggering, flavor tagging, etc.. \cite{lhcapp}.

Recently, different algorithms and techniques based
on Artificial Neural Networks, Genetic Programming and Machine Learning  have been implemented for the
studies as proposed in this paper, namely trying to explain, and modeling of, multiplicity distributions of hadron-nucleus \cite{bakery} and $pp$ interactions \cite{radi1,radi2}.
The motivation to use Artificial Intelligence and Deep Neural Networks (DNN) for such studies is its ability to learn the complex relation between input interaction variables and output observables that arise in $pp$ collisions since such interactions are hard to describe due to the absence of the information on how to describe the quantity of interest with the relevant interaction variables mathematically \cite{wolter}. 

The test we propose is to check if suitable DNNs 
will allow to
predict e.g. the multiplicity distributions at other center of mass energies
than those used in the learning process assuming
and provided no (significant) new physics processes set on in the new energy regime. In the example studied in this paper
we use the multiplicity distributions of charged-particles measured at
energies where LHC collider has collected data and check the ability to predict such distributions for both intermediate new energies and in a new
regime reachable by possible future extensions 
of the CERN $pp$ program such as  a High-Energy LHC Collider (HE-LHC), i.e.  at 27 TeV  that could be located in the present LHC tunnel, based on Future Circular Collider (FCC-hh) magnet technology under development \cite{fcc2}. Furthermore,  we include the proposed   100 TeV  FCC-hh \cite{fcc1}, potentially to  be built using a new 100 km ring circumference. The predictions  are obtained using the LHC data collected at 0.9, 2.36, 2.76, 5, 7, 8 and 13 TeV as input to the model training, i.e. CM energies at which the LHC has operated so far.


The strategy of this study is as follows. 
This study is a 
proof of principle of the underlying idea using 
the PYTHIA event generator instead of real data.
This has the advantage that a uniform analysis can be performed with the "data sets" of all CM
energies and that these are also available to be used as inputs. charged-particle multiplicity distributions are not available yet for all
CM energies.

We set up a machine learning configuration and  train the network with
the $pp$ multiplicity and transverse-momentum distributions of charged-particles  generated using PYTHIA event generator for seven increasingly wider pseudorapidity intervals and for different center of mass energies corresponding to the energies that the LHC operated at till 2018. We use other energy 
settings for those that may be collected in the future to test and support our proposed technique. 
We check the quality of the resulting model's 
ability to predict generator distributions at
different CM energies, including how well these interpolate between the measurements already made and how well they 
can predict distributions for higher energies.

As mentioned, a  practical application for a real world prediction would require to use as input actual measurements based on data. 
At this point in time, these measurements have not been conducted for all CM energies at 
which the LHC was operated. 
Minimum bias charged-particle 
multiplicities distribution measurements do exist, and have been
provided in particular by the CMS and ALICE 
collaborations over the last years.
We hope that studies such as this one will
strongly encourage that such measurements will be  performed and published in future.
Using such a method for predicting higher energies  has the obvious drawback that if a strong new 
physics process will set on in between the region of the measurements and the new energy, this method will 
obviously not make a correct prediction. But turning this argument around: such deviations, when compared with the future data can then point to something new!

This paper has six further sections. Section 2 introduces  the basics of the
DNN. Section 3 gives a summary of our method to collect and preparing data. Section 4 explains in detail our model for predictions. Sections 5 and 6 discuss the results and the conclusion respectively.

\section{Deep Neural Network}



In ML modeling, an approximating function that describes the relation between inputs and outputs can be inferred automatically from the input data without providing explicit information about this function. The most powerful technique to infer an approximation $f(x, w)$ of the unknown function 
$f(x)$ is called  supervised learning, in which the training process contains datasets that conclude inputs and the corresponding targets (desired outputs). The goal of learning is to determine the parameters $w$ of the model, so we can obtain functional approximation for the desired input-output map. In high energy physics, the training data is generally obtained from Monte Carlo simulations  \cite{multi}.

\begin{figure}[h]
\centering
    \includegraphics[width=6cm , height=5cm]{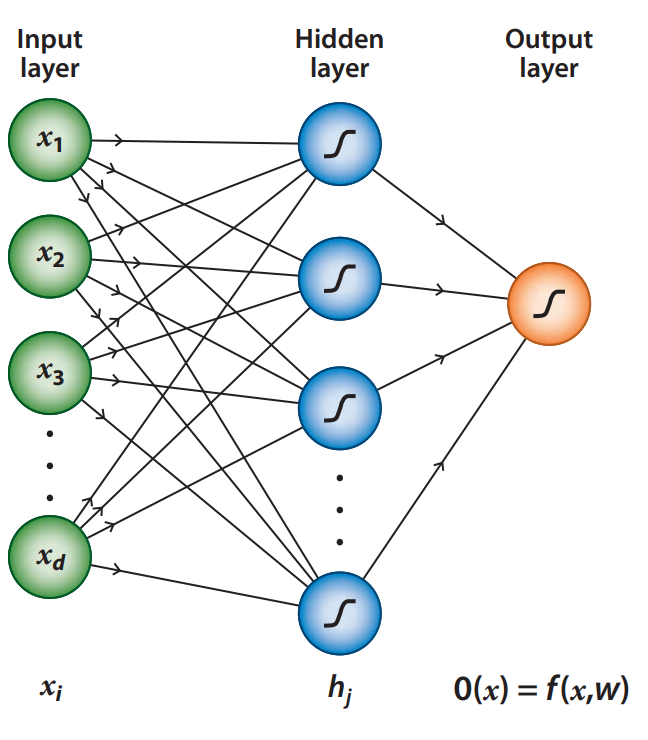}
\caption{A simple feed-forward Neural Network with three layers, from \cite{multi}. }
\label{snnn}
\end{figure}

Feed-forward Neural Networks are the most popular and widely used multivariate methods \cite{multi}. 
It contains an interconnected group of neurons ordered in sequential layers, where each neuron has a role to process the received  information with what is called an activation function, see  
section 
\ref{sec3}, then the result is moved to the next layer of nodes. The first layer, which receives the input variables is called the input layer, followed by one or more hidden layers. The last layer is responsible for the final response of the neural network and is  called the output layer. Each interconnection is specified by weight and bias, which are the network parameters that are
being learned and updated during the training process. A simple NN is shown in Fig. \ref{snnn}.

In Fig \ref{snnn}, showing a NN that consists of one hidden layer of nodes and an input data layer with $d$ feature variables (inputs)
$x = \{x_{1} , x_{2} , . . . x_{d} \}$, the output of this network is
\begin{equation}
\centering
    f(x,w)= g(\theta + \sum_{j}^{} w_{j}b_{j})
\end{equation}

where $g$ represents the activation function and $b_{j}$ is the output from the hidden neurons:

\begin{equation}
\centering
    b_{j}= g(\theta_j + \sum_{j}^{} w_{ij}x_{i})
\end{equation}

\section{Data preparation}
\label{data}
PYTHIA \cite{PythiaPhysicManual} is a general-purpose Monte Carlo event generator that is actively  used in particle physics, both in general and for the LHC in particular. This generator has undergone decades of development and tuning to colliders and other data.

The event generation  consists of several steps starting typically from a hard scattering process, followed by initial- and final-state parton showering, multi-parton interactions, and  the 
 final hadronization process. PYTHIA uses different model approaches for these steps, e.g. it uses a $p_{T}$-ordered perturbative approach \cite{pythiapartonshowering} for modeling of parton shower. The original impact parameter model \cite{mulitpleinteractionpythia} for multi-parton scattering and the  Lund string fragmentation model \cite{pythiahadronization1,pythiahadronization2}
 are used for the hadronization (fragmentation) of partons into 
 hadrons.
 
The proton-proton collisions are generated in this work with the PYTHIA 8.186  \cite{pythia} version of the program. The collisions are generated at  $\sqrt{s} = 0.9, 2.36, 2.76$, $5$, $7$, $8$, and $13$ TeV, corresponding to the energies at which the LHC was operated from 2010 till 2018, in order to train and evaluate the model performance, and 
at the energies  $\sqrt{s} = 27, 50, 100, 150$ TeV in order to compare with the prediction of our model and to show its ability to predict event distributions at the energies that were not used
to train on. Different model response
functions are extracted for different pseudorapidity intervals. In total 50*$10^6$ collisions were simulated at 7, 8 and 13 TeV, and 5*$10^6$ events were generated for other CM energy values, using default minimum bias generation settings of the generator, discussed below. 
The difference in the number of events was chosen to emulate the experimental situation where much 
larger data sets were collected  at 7, 8 and 13 TeV at the LHC, than for the other CM energies.

The inelastic (diffractive and non-diffractive) proton-proton collisions were simulated using the PYTHIA Monash 2013 tune \cite{monsh}. The Monash parameters are tuned such that these  provide a reasonable description of the experimental data at LHC energies for the bulk of the minimum bias charged multiplicity distribution and several other event characteristics.

Minimum bias events and particles are selected 
in this study
according to the following criteria. Each event must have at least one charged-particle in the final state which is emitted within the studied pseudorapidity interval  and within the full acceptance of the azimuthal angle  ($\phi$), 
and with a minimum $p_{T}$ $>$ of 400 MeV.
The number of events that pass those selection criteria at the different energies and pseudorapidity intervals are given in Table \ref{tab1}.

\begin{table}[h]
\caption{The number of events that pass the selection criteria at different energies and different pseudorapidity intervals.}
\label{tab1}
\centering

\begin{tabular}{llllllll}

\hline
\multirow{2}{*}{$\sqrt{s}$} & \multicolumn{7}{c}{ The number of events at different $\Delta\eta$ intervals (*$10^6$)} \\
 & \multicolumn{1}{c}{0.5}       &  \multicolumn{1}{c}{1}      &   \multicolumn{1}{c}{1.5}     &    \multicolumn{1}{c}{2}    &  \multicolumn{1}{c}{2.5}     &   \multicolumn{1}{c}{3}    &   \multicolumn{1}{c}{3.5}   \\
\hline
\multicolumn{8}{c}{$p_{T}>400$ MeV} \\
\hline

0.9  &     2.9221   &      3.6344  &     3.9446   &    4.1190    &   4.2369    &   4.3264    &4.3994      \\
2.36 &    3.1444    &   3.7824     &    4.0500    &   4.1987     &   4.2999    &    4.3785   & 4.4435     \\
2.76 &     3.1794   &    3.8044    &   4.0659     &    4.2105    &   4.3092    &   4.3854    & 4.4490     \\
5   &   3.3099   & 3.8861    &   4.1253   &   4.2576    &     4.3471   & 4.4163   & 4.4745     \\
7 (5m)    &   3.3748    & 3.9275     &    4.1567    &   4.2822    &  4.3673  &  4.4334    &  4.4893     \\
7 (50m)    &   34.572    &  39.890     &    41.942    &   43.038    &    43.792   &  44.399    &  44.928     \\
8 (5m)   &  3.4016    &  3.9440   &  4.1688  &  4.2917     & 4.3753    &   4.4402   & 4.4951    \\
8 (50m)   &   34.818    &  40.049    &   42.064   &  43.137     & 43.874    &   44.469    & 44.987     \\
10   &  3.4432   &3.9710    &  4.1891     &   4.3084    &  4.3892   &  4.4520  & 4.5049    \\
13 (5m)   &   3.4908     & 4.0017     &  4.2121      &   4.3270    &   4.4051  &  4.4659  & 4.5169     \\
13 (50m)   &   35.680     & 40.596     &  42.481      &   43.483    &   44.170   &  44.720  & 45.198     \\
20   &    3.6374    &  4.1042      &  4.2830      &   4.3775     &   4.4423    &    4.4940   & 4.5382     \\
27   &  3.6836  &  4.1338    & 4.3053&   4.3965   &    4.4585   & 4.5081 & 4.5507    \\
50   &    3.7727    &  4.1911      &    4.3513    &   4.4357     &    4.4932   &   4.5388    &  4.5777    \\
100  &    3.8611    &    4.2502    &    4.3987    &    4.4773    &    4.5299   &    4.5716   &  4.6069    \\
150  &   3.9089   &   4.2819   &     4.4239    &     4.4995   &   4.5502   &   4.5897  &  4.6234  \\




       \hline
\end{tabular}

\end{table}

\section{Prediction Network}
\label{sec3}

The software package used in this study for the modeling is Keras \cite{keras} version 2.4.3, which is an Open Source Library for Neural Network written in Python version 3.8.6 and built on top of TensorFlow \cite{tensor} version 2.4.1. The importance of this tools is reducing the role of the physicist to choose an appropriate problem, data scaling and manipulation, DNN architecture, and training technique.

\begin{figure}[h]
 \begin{subfigure}{0.55\textwidth}
    \includegraphics[width=\textwidth , height=6.5cm]{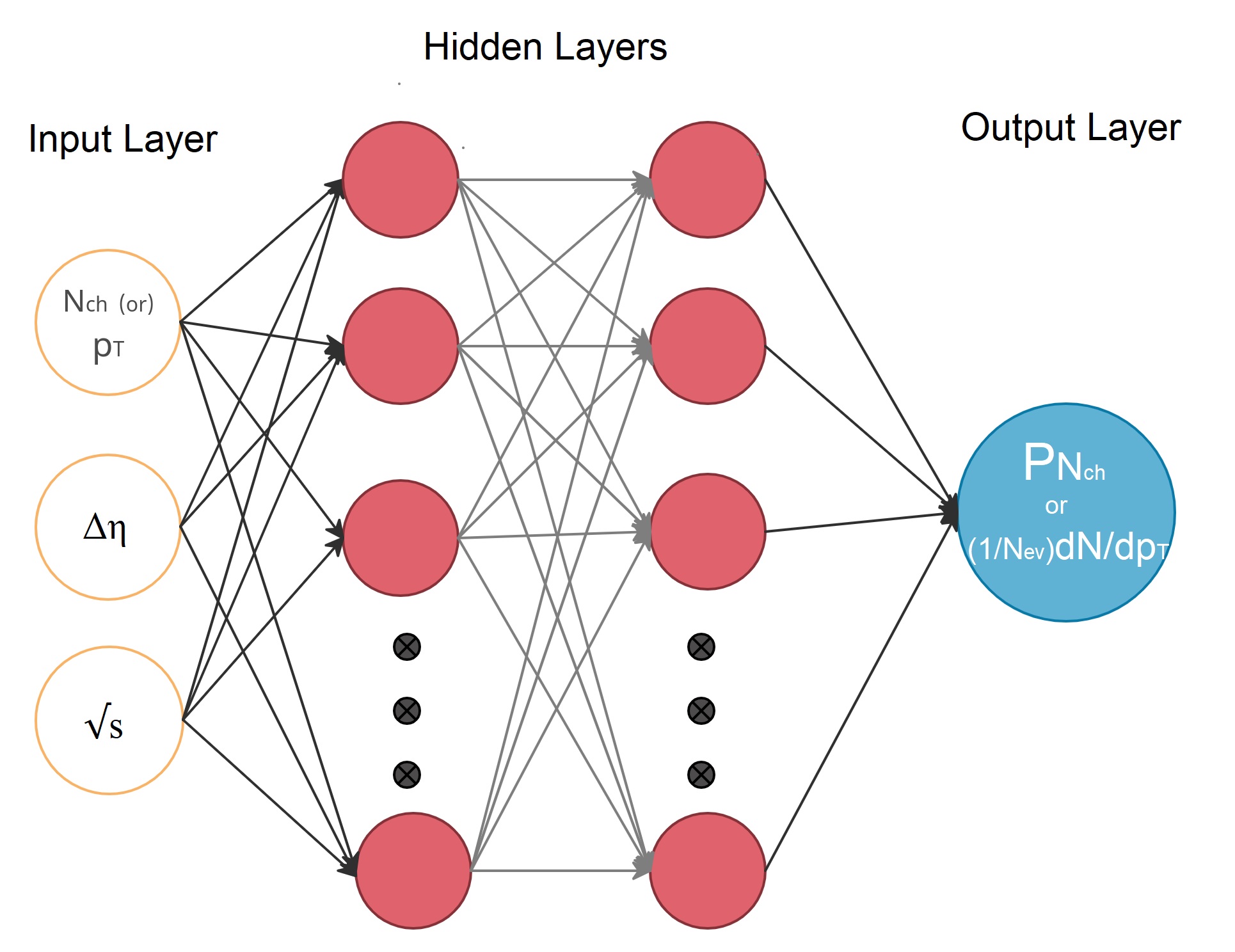}
\caption{ } 
\end{subfigure}
  \hfill
   \begin{subfigure}{0.45\textwidth}
    \includegraphics[width=\textwidth , height=6.5cm]{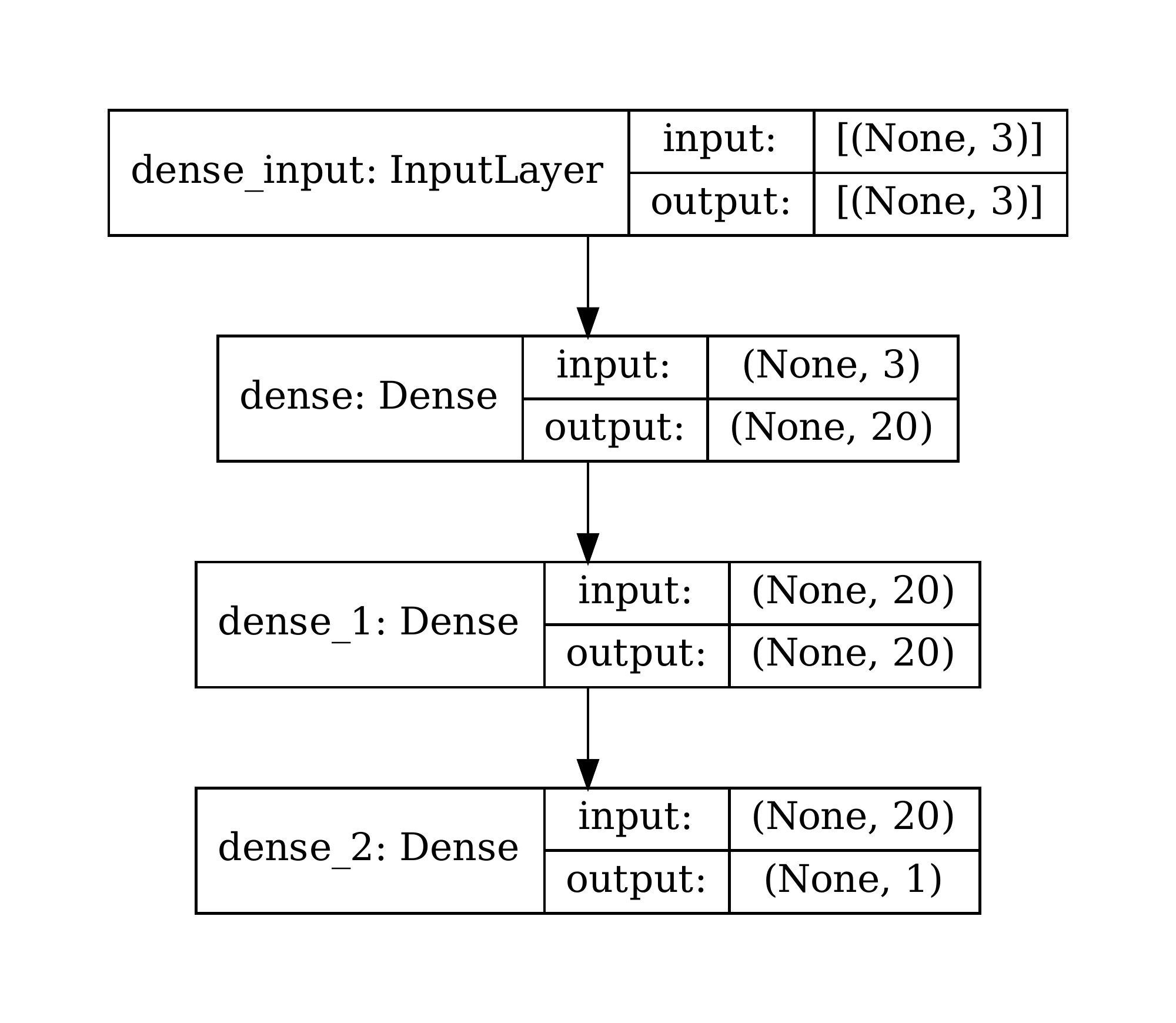}
\caption{ } 
\end{subfigure}
\caption{(a) is a schematic diagram for our proposed neural network. (b) is a plot representing the number of layers and number of inputs and outputs for every layer.}
\label{net}
\end{figure}

Several DNNs were tried to
solve the problem, with varying number of internal
layers and neurons per layer. The DNN model that showed a very good agreement between the probability $P(N_{ch},\sqrt{s},\Delta\eta)$ and the charged-particle multiplicity ($N_{ch}$) at different pseudorapidity windows ($\Delta\eta$) and different collision energies ($\sqrt{s}$) consists of an input layer with three inputs, two hidden layers with each 20 neurons and final output layer with only one output, see Fig. \ref{net}, was chosen for this study. This model shows also an excellent agreement for the transverse-momentum ($p_{T}$) distributions but with input ($p_{T}$, $\Delta\eta$, $\sqrt{s}$) and the output the model trained on  $(1/N_{ev})dN/dp_T$ which is the distribution giving the number of particles as 
function of $p_{T}$, divided by the number of events which have at least one particle 
 with $p_{T}>$ 400 MeV within the studied rapidity range.

The initial random weights and biases of the Keras layers are set using the "kernel\_initializer" and "bias\_initializer" to follow a normal distribution. The activation function implemented for the hidden layers is a hyperbolic tangent "tanh" \cite{tanh,linear}, namely $f(x)= \frac{\sinh(x)}{\cosh(x)}=\frac{e^{x}-e^{-x}}{e^{x}+e^{-x}}$, a nonlinear function to allow for a flexible modeling and the output ranges from $-1$ to $1$. Furthermore, the activation function for the output layer is "linear" \cite{linear}, namely $f(x)=x$.
The role of the activation function is to analyze the total information received by the neuron and this determines the output information produced by 
the neuron in response to the input information.

The loss value, which quantifies the amount of information lost, used in this model is the mean absolute error (mae) between the true value and the predicted one. Mathematically, if $\gamma$ is a vector of $n$ predictions, and $Y$ is the vector of $n$ observed values, then:


\begin{equation}
mae= \frac{1}{n}\sum_{i=1}^{n}|\gamma - Y|\end{equation}

The optimizer used for this model is the "Adam"\cite{adam} optimizer with a 0.0005 learning rate. 
This optimizer is used for improving the speed and performance of the training of our model. 

We further set the model  "batch\_size"=100 and in order to overcome over-training, we have used the EarlyStopping class \cite{earlystop} with min\_delta=$e^{-5}$ and "patience" = 1000 in order to stop the processing after the model has reached the smallest loss value for the validation data.


The $pp$ collisions generated by PYTHIA at 0.9, 2.36, 2.76, 5,  7 , 8  and 13  TeV are separated into two parts. Two third of the data is used for model training, and the other one-third is used for model validation. The number of events at those energies and different pseudorapidity windows are presented in Table~\ref{tab1} (for the  transverse model only 
5m data sets are used while for the multiplicity studies the 50m data sets were included).

The best prediction results are obtained when training the multiplicity model with 67\% of 0.9, 2.36, 2.76, 5, 7 , 8  and 13  but in case of the transverse-momentum a better training was
achieved, with less bias, using training samples
based on the same statistics and hence the samples
with 5m collisions each at the different energies
were used for this study.

The input values that are used to train the multiplicity model used are $N_{ch}$*0.1, $\sqrt{s}$ and $\Delta\eta$ and the output is $P(N_{ch},\sqrt{s},\Delta\eta)$. Empirically
we found that using a reduced value range for 
$N_{ch}$ lead to more stable and lower bias
results, as it keeps the range of inputs close to each other, so there is no input intrinsically influence the model due to its large value.
The multiplicity and $p_T$ distributions cover several orders of magnitude in the bin population, hence for a more
stable training procedure and in order to avoid large biases, the training is performed using 
the logarithms if the bin values for both
studies.
Furthermore, the number of events with a specific multiplicity must be larger that 10 in order to remove any fluctuations in the spectrum tails.


The TensorFlow random seed values are set to one at the start, then deploy training  until it reaches the value of the smallest loss value compared to validation data, then the weights and biases that give the least loss are taken. For the comparisons, the results are shown using the original un-scaled values and will be discussed in the next section.

Next, the model is used to predict the energies at future collider energies, e.g. for 
an upgraded LHC to run 
at higher energy, i.e. 20 TeV 
and 27 TeV. Furthermore, this model can be 
tested for predictions for much higher 
energies, as expected at the Future Circular Collider (FCC)  i.e. 100 TeV. We also test the 
predictive power for the highest imaginable energy
to date for a 100 km ring if  the technology would 
allow for producing 24T instead of 16T magnets
superconducting magnets, which would lead to 
collisions at 150 TeV. Such ideas have been 
mentioned as a possible --but yet to be demonstrated
-- upgrade option beyond the baseline for the 
SPPC machine in the Chinese future collider project proposal\cite{SPPC}.


\section{Results \& Discussion}
\begin{figure}[h]
 \begin{subfigure}{0.5\textwidth}
    \includegraphics[width=\textwidth , height=7cm]{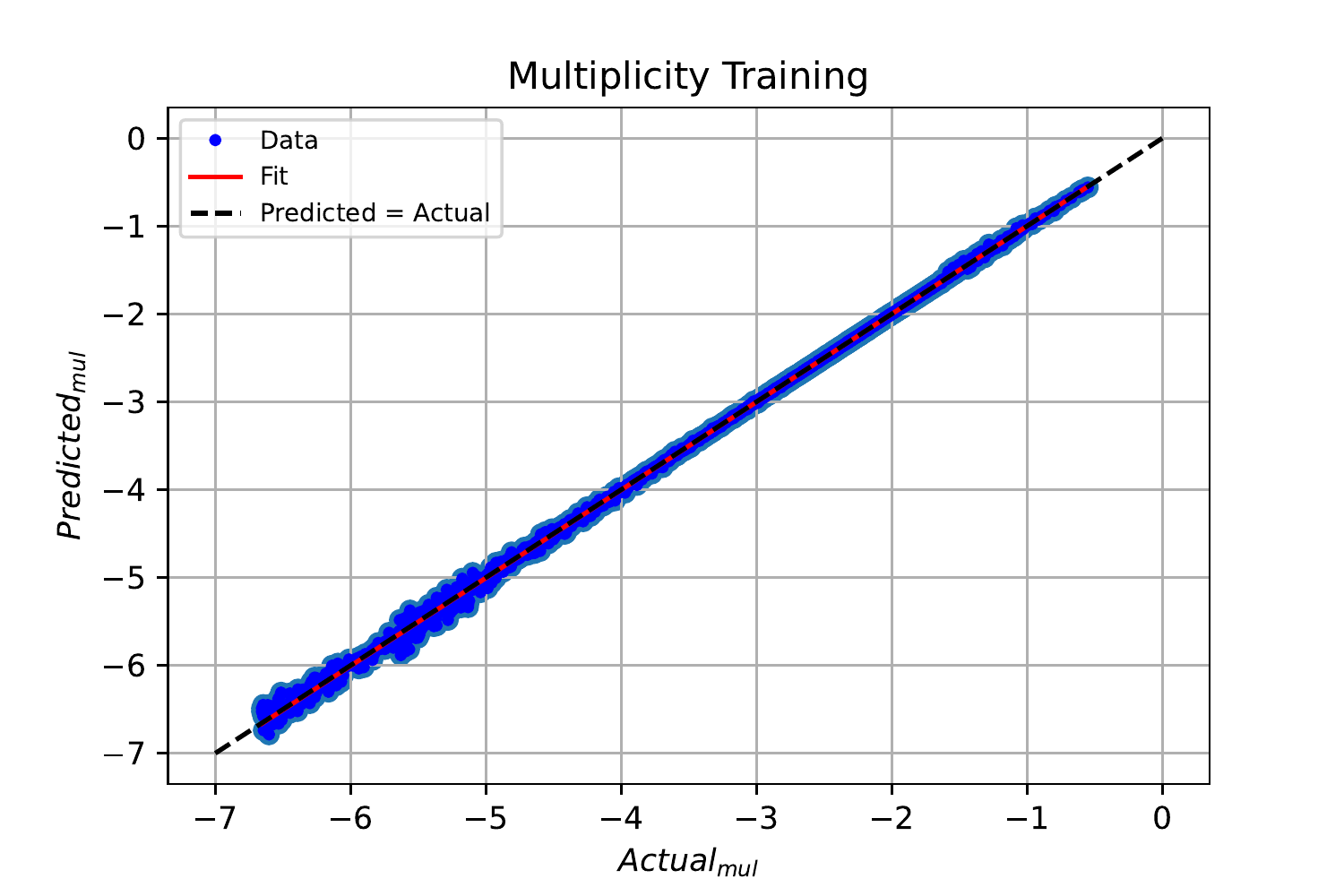}
\caption{for multiplicity. } 
\end{subfigure}
  \hfill
   \begin{subfigure}{0.5\textwidth}
    \includegraphics[width=\textwidth , height=7cm]{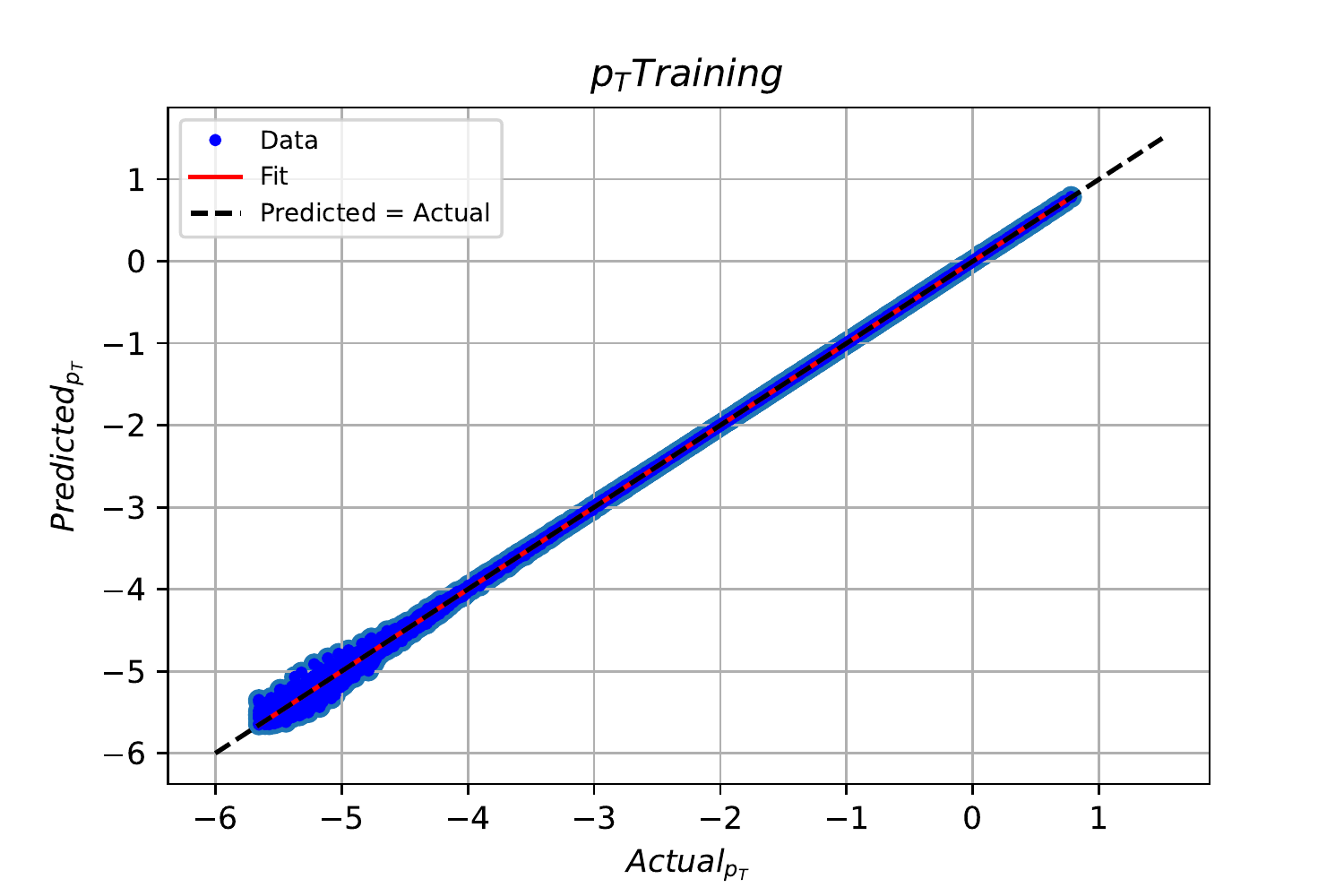}
\caption{for transverse-momentum. } 
\end{subfigure}
\caption{The relation between the predicted and actual output for the training data.}
\label{perform1}
\end{figure}

\begin{figure}[h]
 \begin{subfigure}{0.5\textwidth}
    \includegraphics[width=\textwidth , height=7cm]{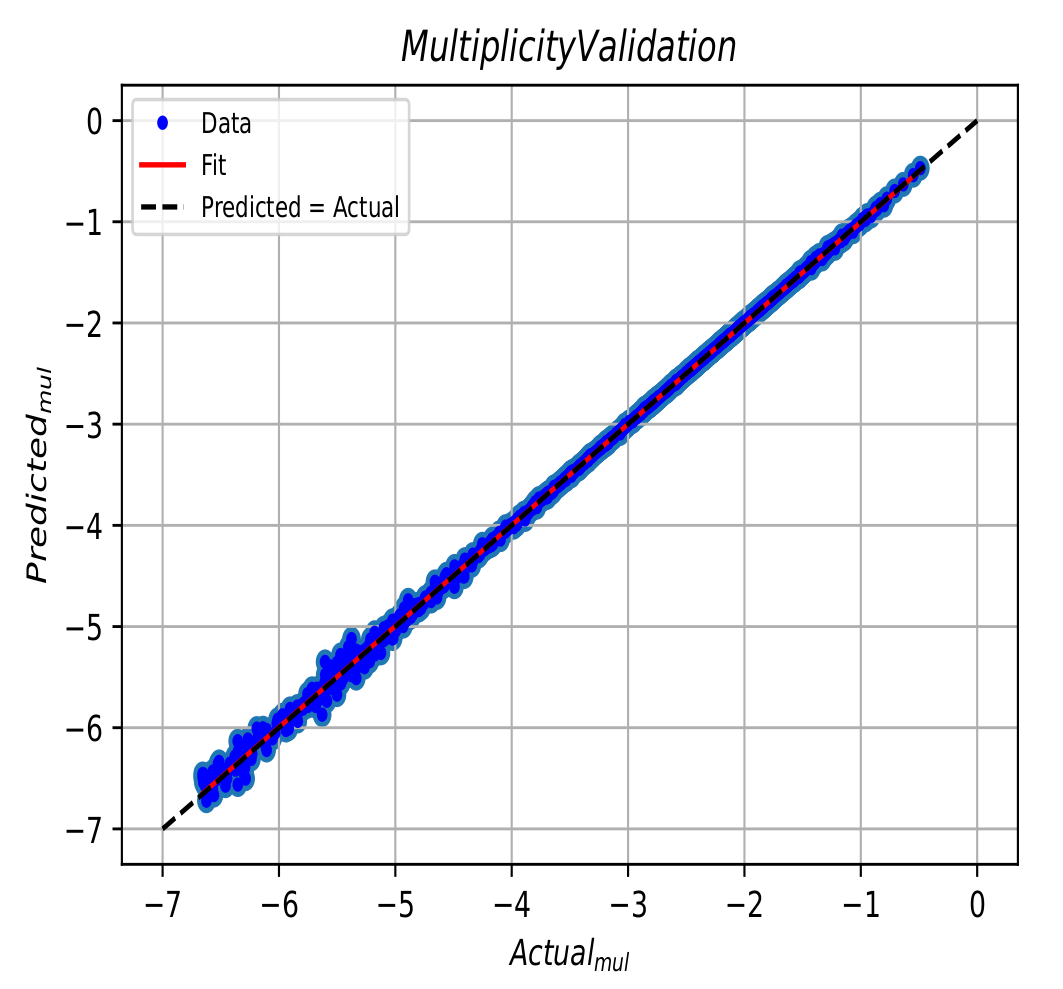}
\caption{for multiplicity. } 
\end{subfigure}
  \hfill
   \begin{subfigure}{0.5\textwidth}
    \includegraphics[width=\textwidth , height=7cm]{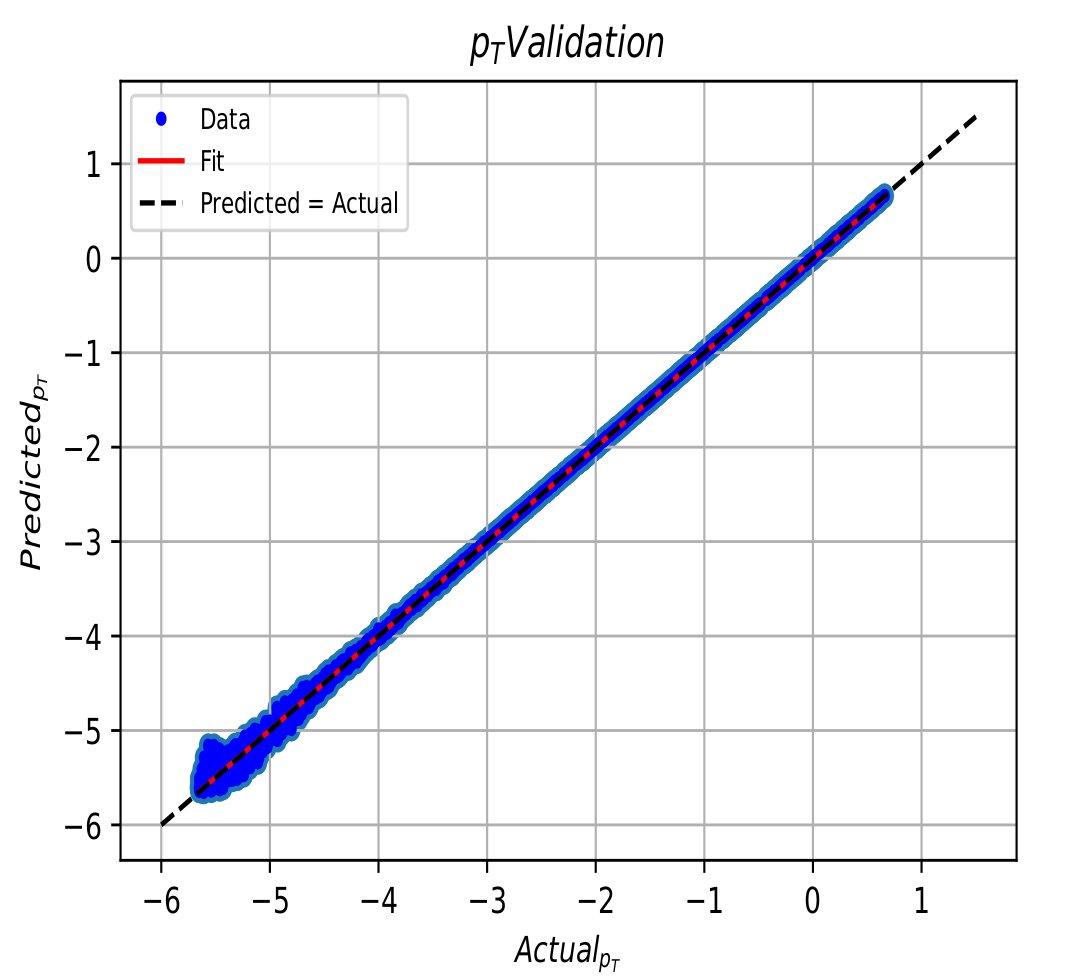}
\caption{for transverse-momentum. } 
\end{subfigure}
\caption{The relation between the predicted and actual output for the validation data.}
\label{perform3}
\end{figure}

The performance of the model is found to be excellent for the  multiplicity and transverse-momentum distributions, as demonstrated by the relation between the true output from PYTHIA and the one predicted by the model for training data in Fig. \ref{perform1} and  in Fig. \ref{perform3} for validation data, both shown on a  logarithmic scale. 


\begin{figure}[h]
 \begin{subfigure}{0.5\textwidth}
    \includegraphics[width=\textwidth , height=7cm]{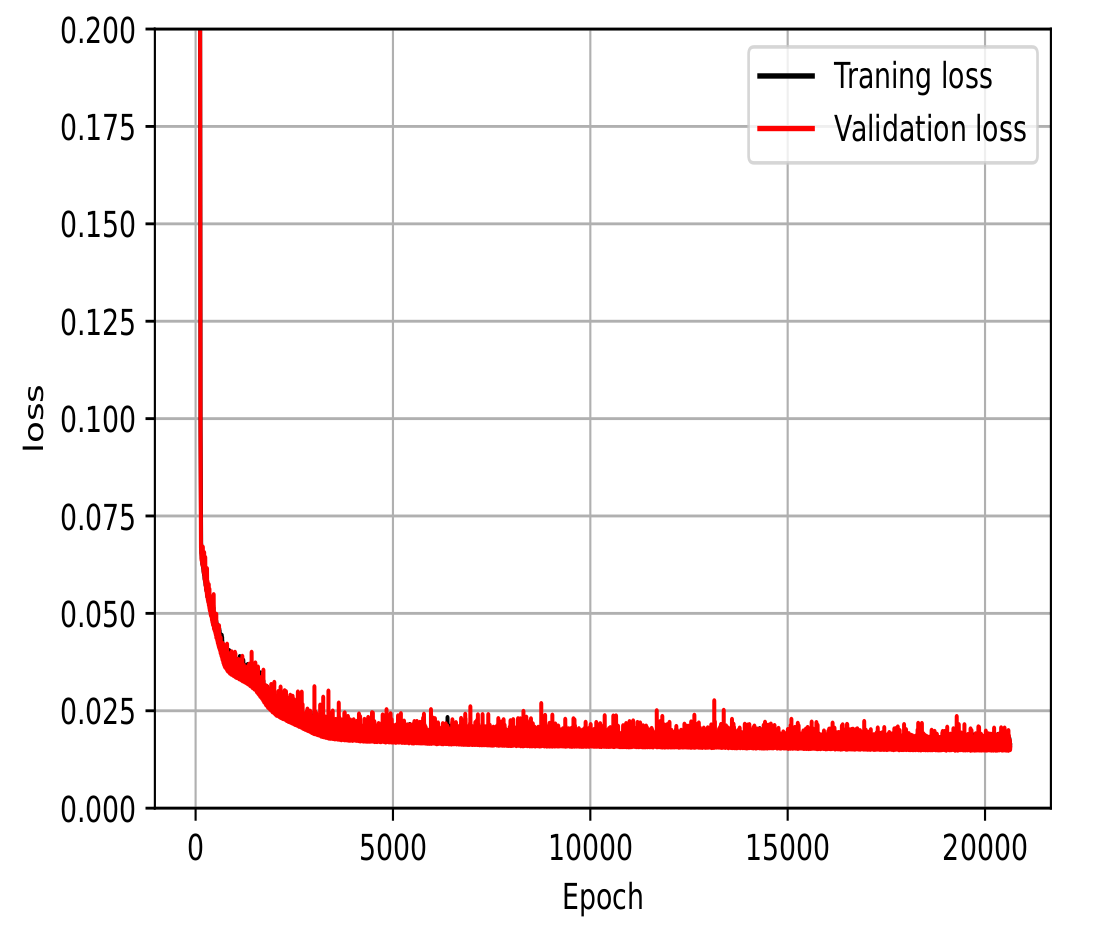}
\caption{for multiplicity. } 
\end{subfigure}
  \hfill
   \begin{subfigure}{0.5\textwidth}
    \includegraphics[width=\textwidth , height=7cm]{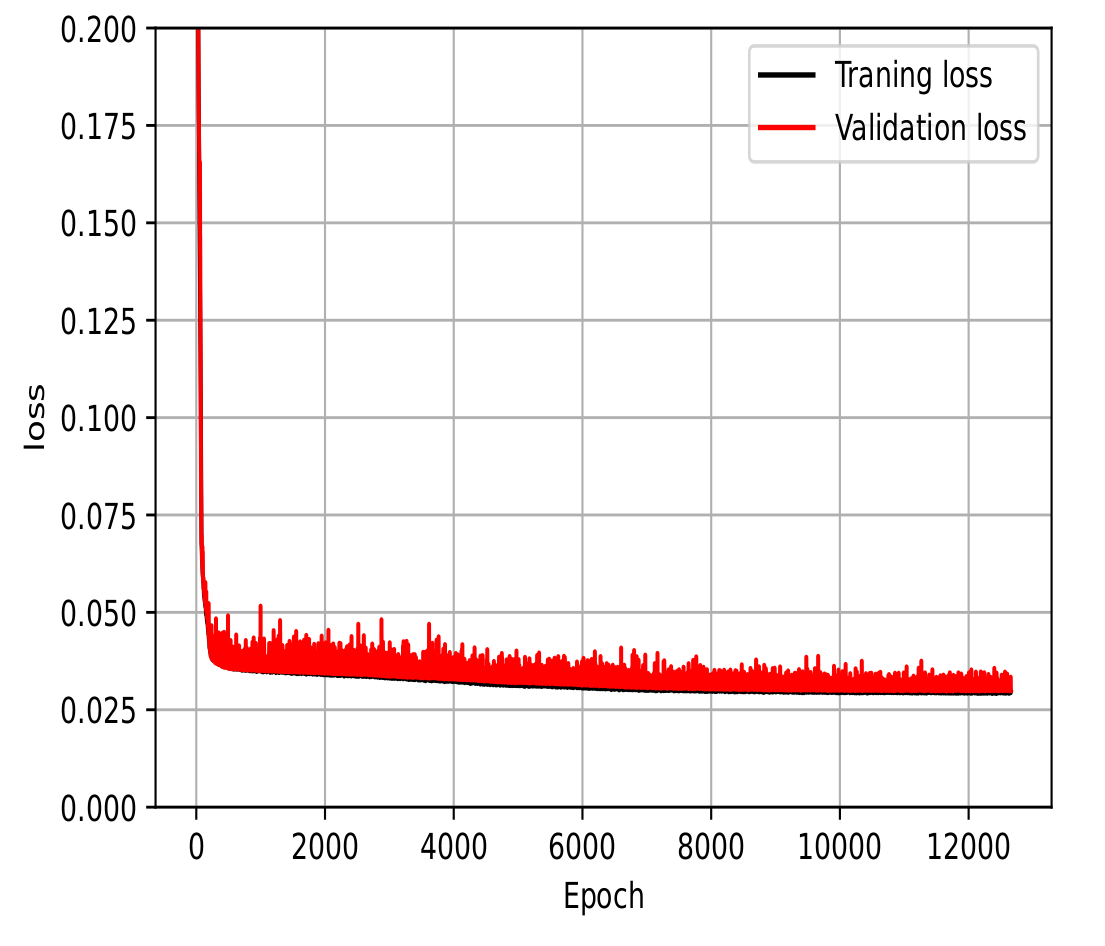}
\caption{for transverse-momentum. } 
\end{subfigure}
\caption{The model training and validation data loss value.}
\label{perform2}
\end{figure}

Fits to a linear dependence are made.
The  result for the training data on the multiplicity gives "$Predicted_{mul}$ = $ 1.0013 * Actual_{mul} +0.0018$" and for the transverse-momentum "$Predicted_{p_{T}}$ = $0.9990 * Actual_{p_{T}} -0.0031$" with $R^{2}$= 0.9995 and 0.9994 in case of multiplicity and transverse-momentum respectively, where $R^{2}$ is the so called coefficient of determination \cite{prob}, which is a measure of the quality of fitting, and defined by 

\begin{equation}
R^{2}=\frac{\sum (\hat{y_{i}}-\bar{y})^2}{\sum(y_{i}-\bar{y})^2}
\end{equation}

where, $y_i$  is the true value, $\hat{y_{i}}$ is the predicted value by the model and $\bar{y}$ is the mean value of all $y_i$ values.

The parameters obtained for the validation data on the multiplicity are "$Predicted_{mul}$ = $0.9983 * Actual_{mul} -0.0047$" and  transverse-momentum are "$Predicted_{p_{T}}$ = $0.9985 * Actual_{p_{T}} -0.0042$" with $R^{2}$= 0.9995, 0.9990  respectively.
These values show that there is very little bias.

Another important and recommended test of the model quality is shown in  Fig. \ref{perform2} as the loss value of the training and the validation data is almost the same which demonstrates that this model doesn't suffer from under/over fitting. 

\begin{figure}[!tbp]
 \begin{subfigure}{0.52\textwidth}
    \includegraphics[width=\textwidth , height=6cm]{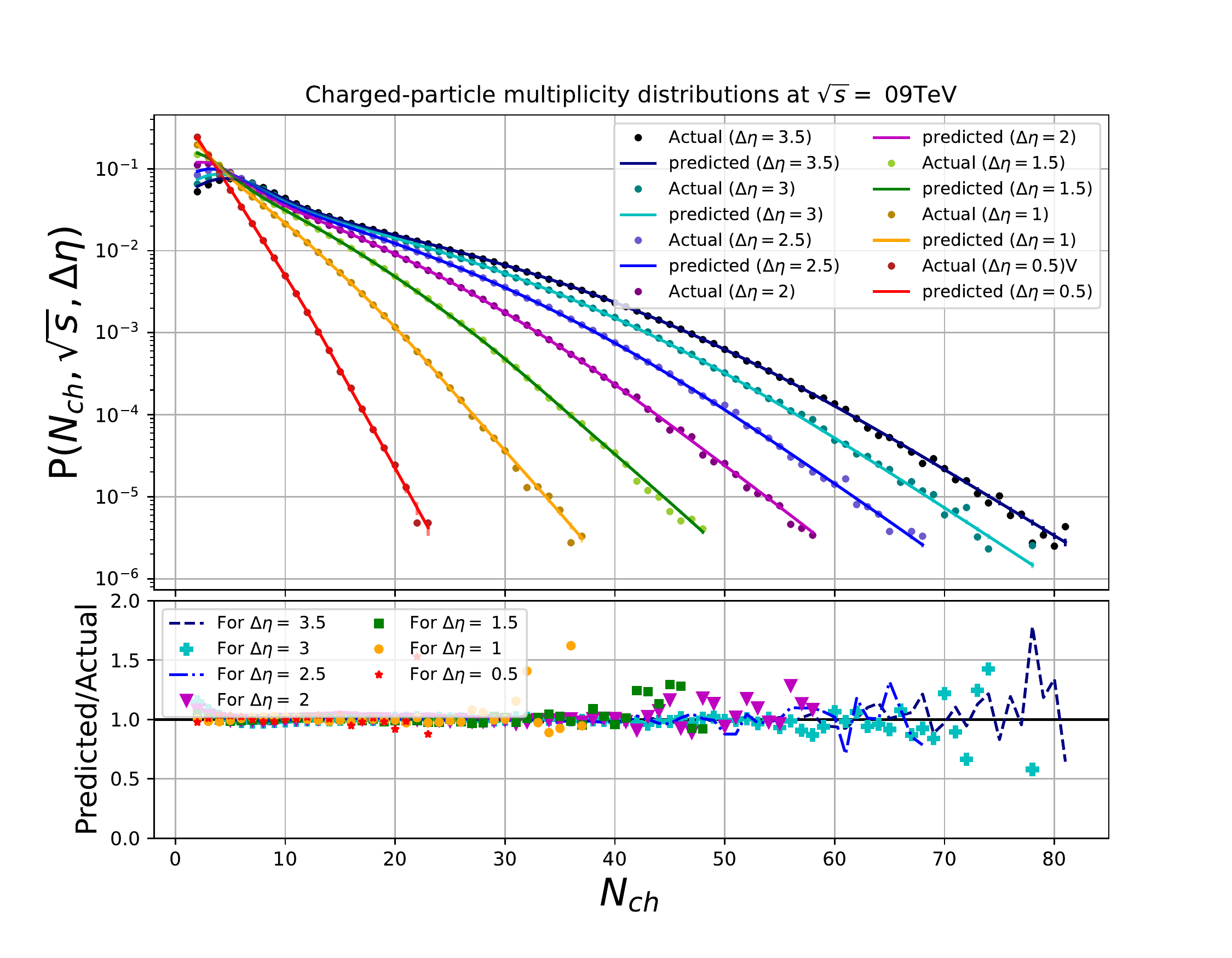}
\caption{\footnotesize for $\sqrt{s} = 0.9TeV$. } 
\end{subfigure}
  \hfill
   \begin{subfigure}{0.52\textwidth}
    \includegraphics[width=\textwidth , height=6cm]{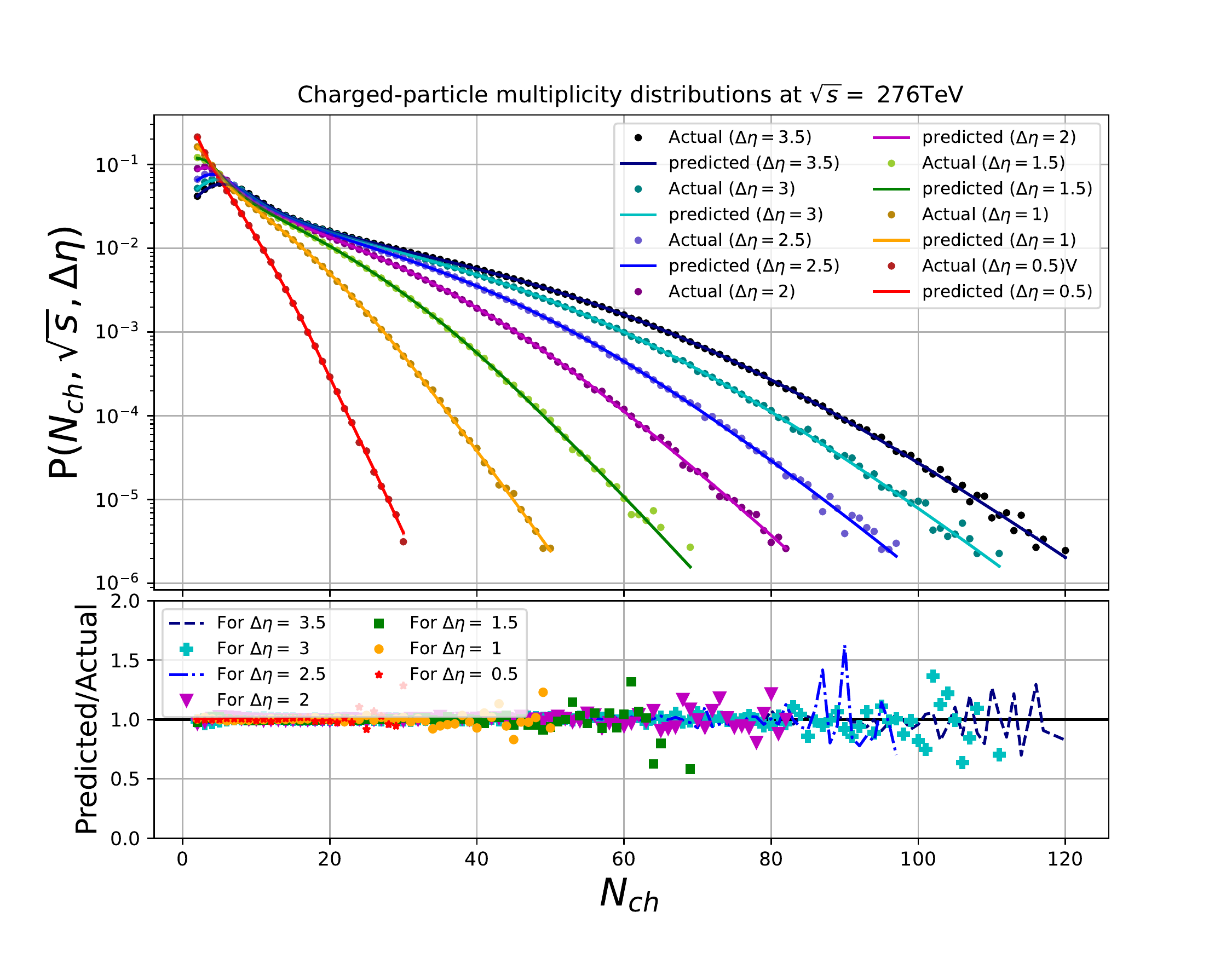}
\caption{\footnotesize  for $\sqrt{s} = 2.76TeV$. } 
\end{subfigure}

 \begin{subfigure}{0.52\textwidth}
    \includegraphics[width=\textwidth , height=6cm]{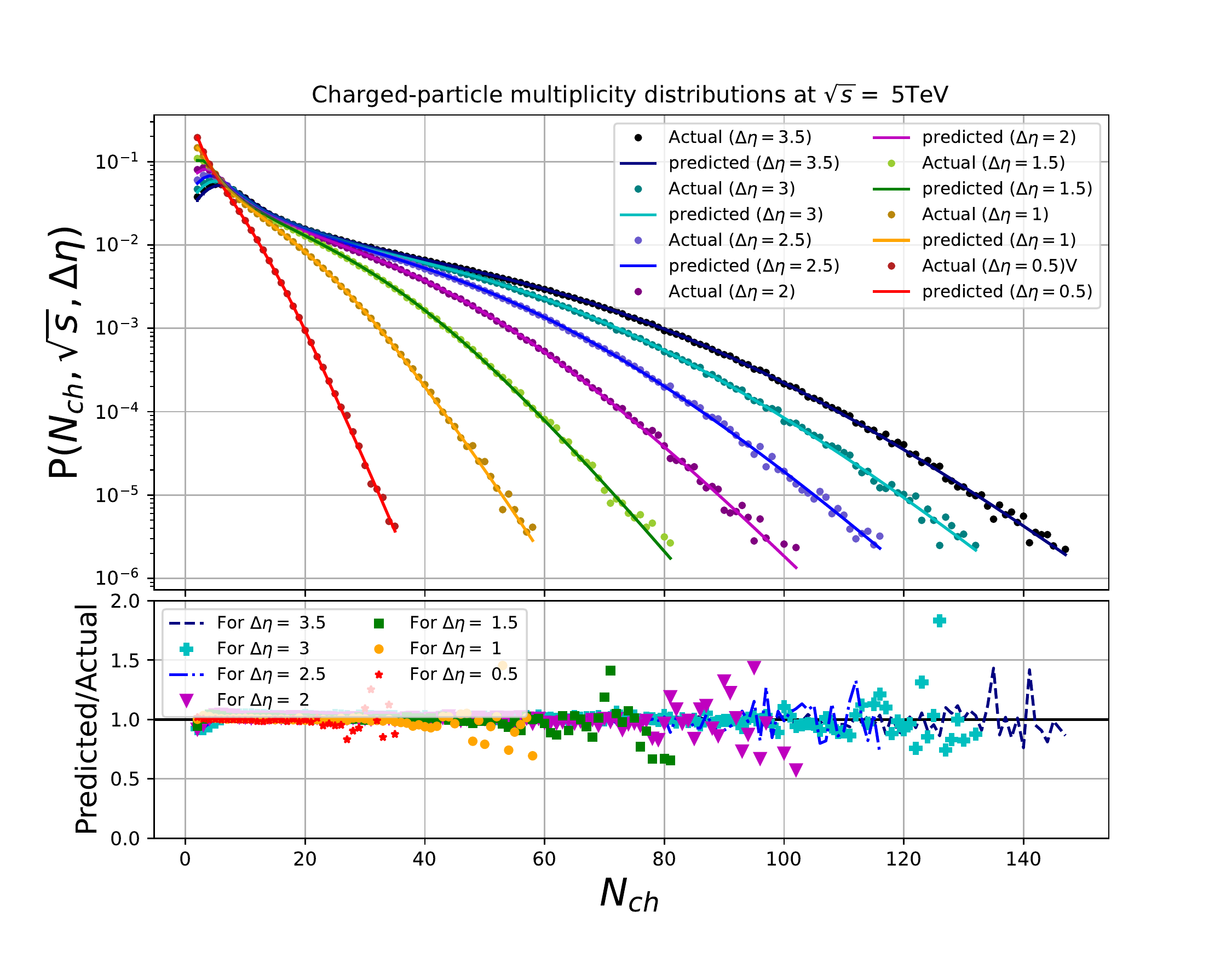}
\caption{\footnotesize for $\sqrt{s} = 5TeV$. } 
\end{subfigure}
  \hfill
   \begin{subfigure}{0.52\textwidth}
    \includegraphics[width=\textwidth , height=6cm]{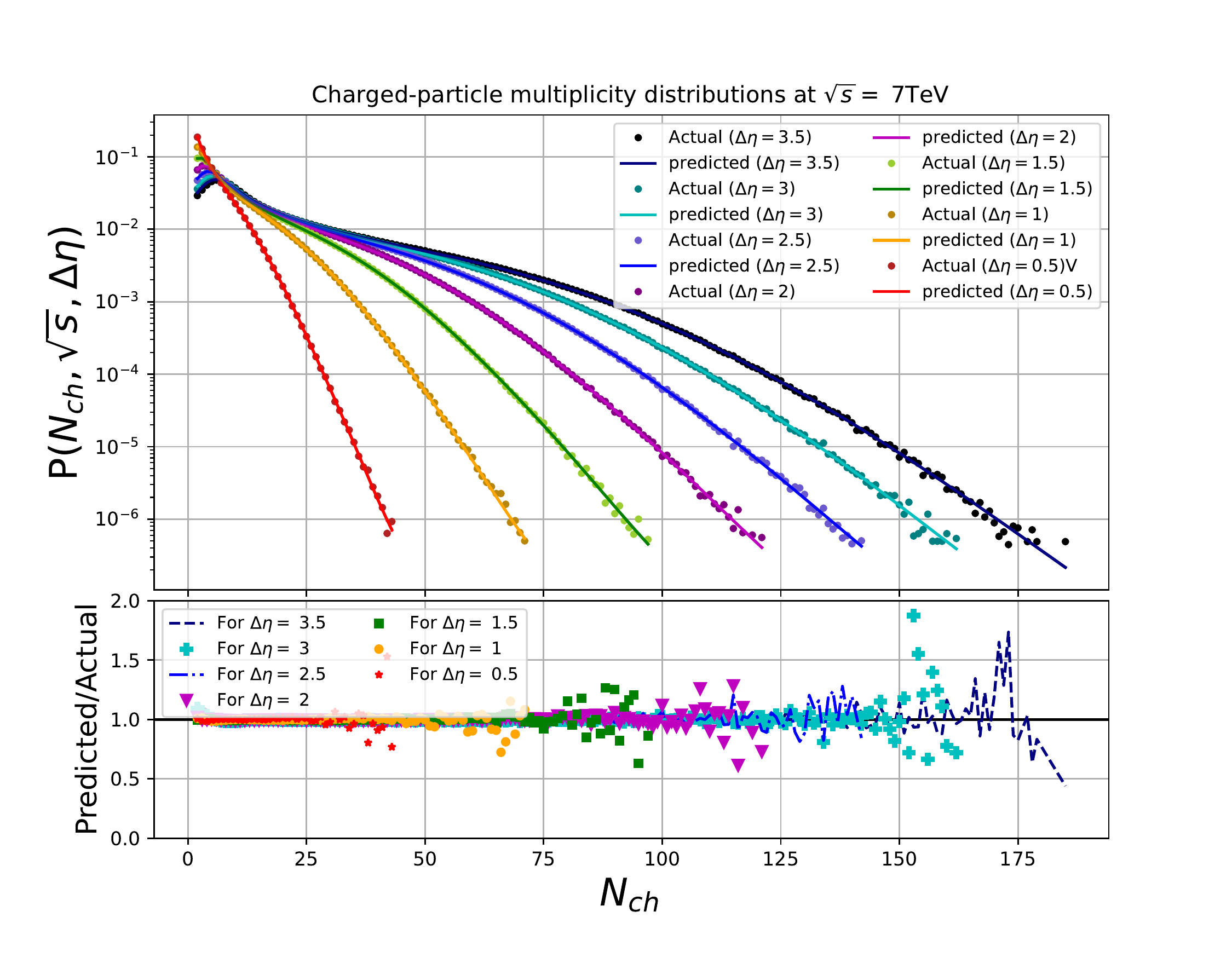}
\caption{\footnotesize for $\sqrt{s} = 7TeV$. } 
\end{subfigure}

 \begin{subfigure}{0.52\textwidth}
    \includegraphics[width=\textwidth , height=6cm]{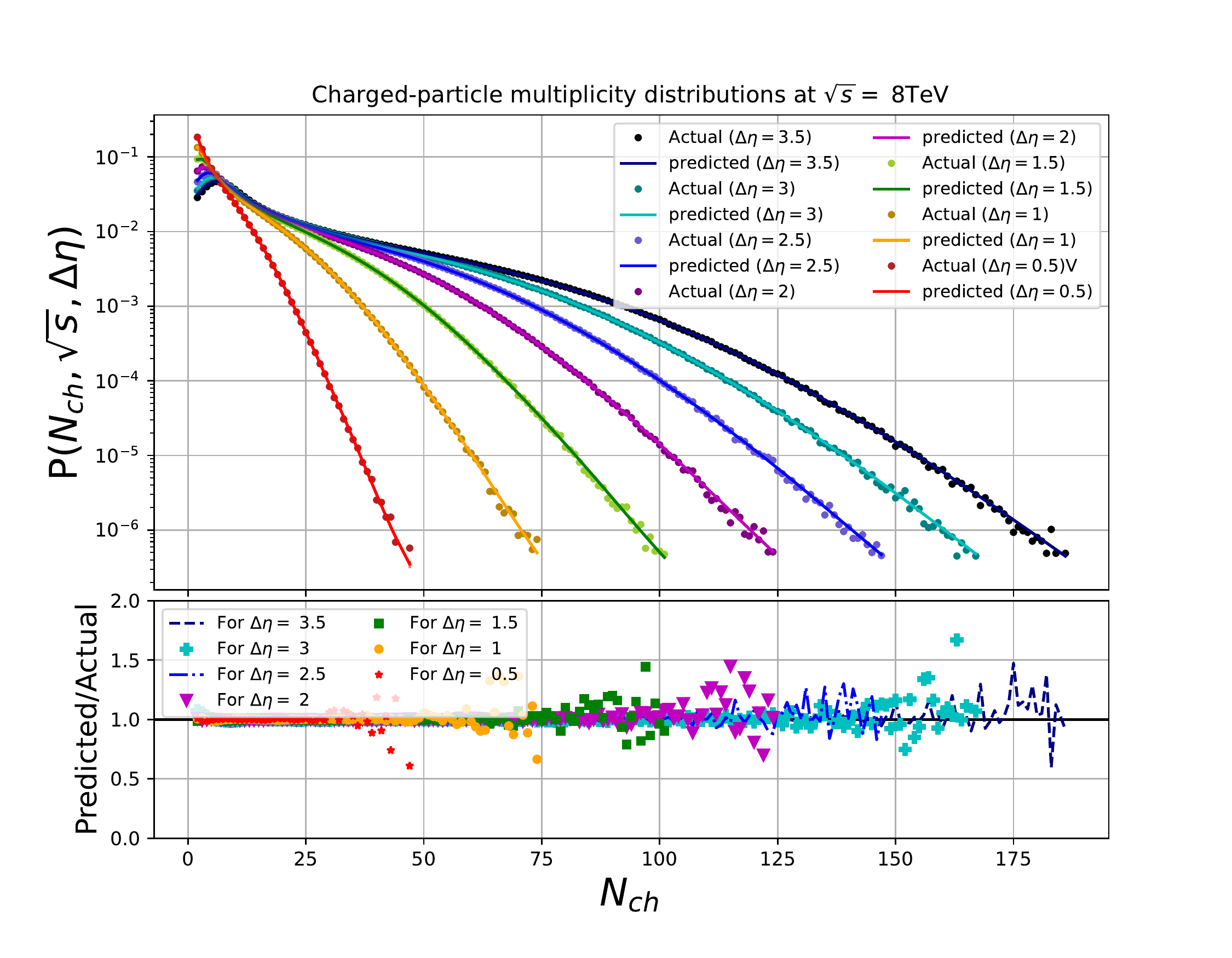}
\caption{\footnotesize for $\sqrt{s} = 8TeV$. } 
\end{subfigure}
  \hfill
   \begin{subfigure}{0.52\textwidth}
    \includegraphics[width=\textwidth , height=6cm]{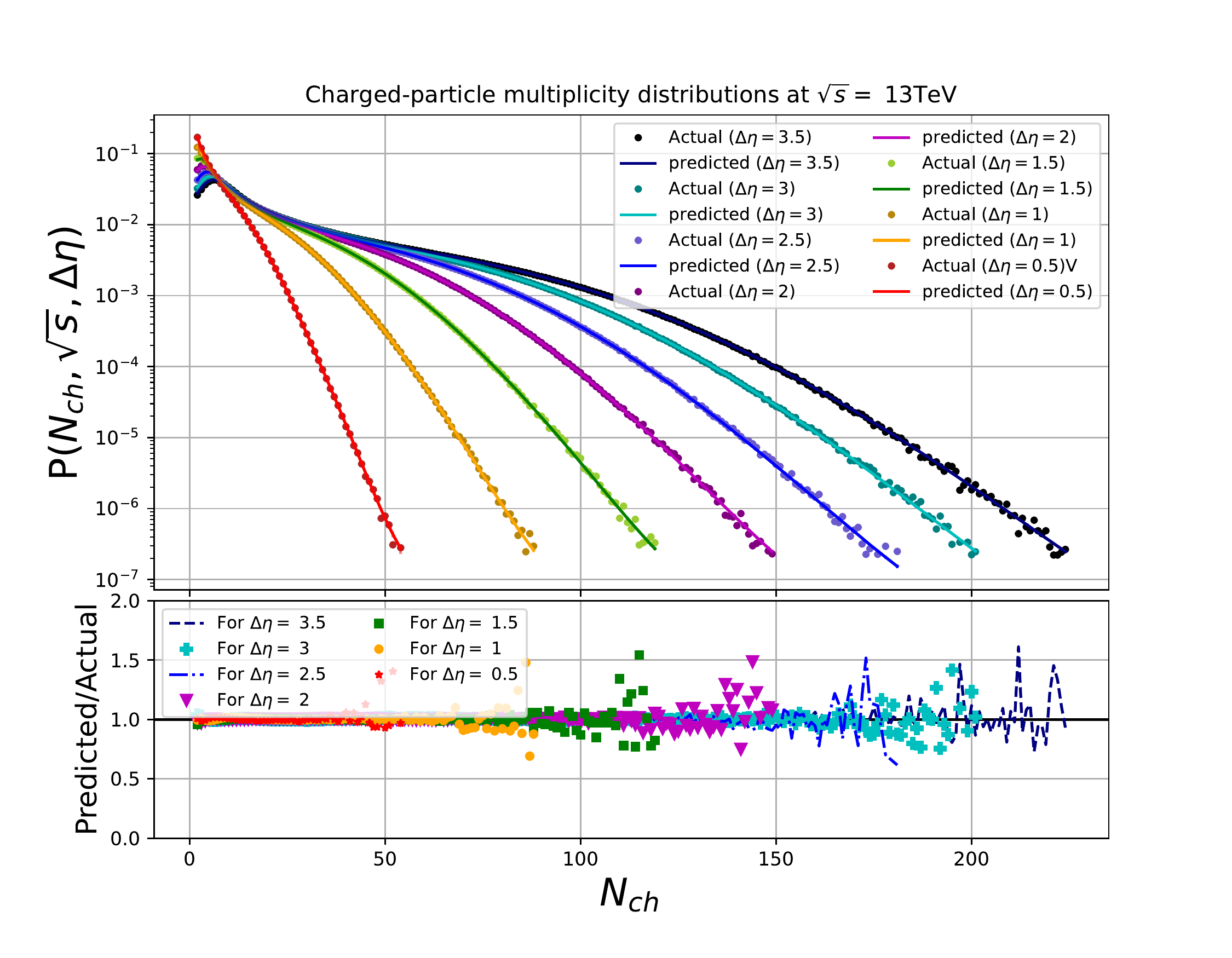}
\caption{\footnotesize for $\sqrt{s} = 13TeV$. } 
\end{subfigure}

\caption{ 
The DNN results in comparison with multiplicity distribution generated by PYTHIA at the training runs (0.9, 2.76, 5, 7, 8 and 13 TeV).
} \label{train}
\end{figure}

\begin{figure}[!tbp]
 \begin{subfigure}{0.52\textwidth}
    \includegraphics[width=\textwidth , height=9cm]{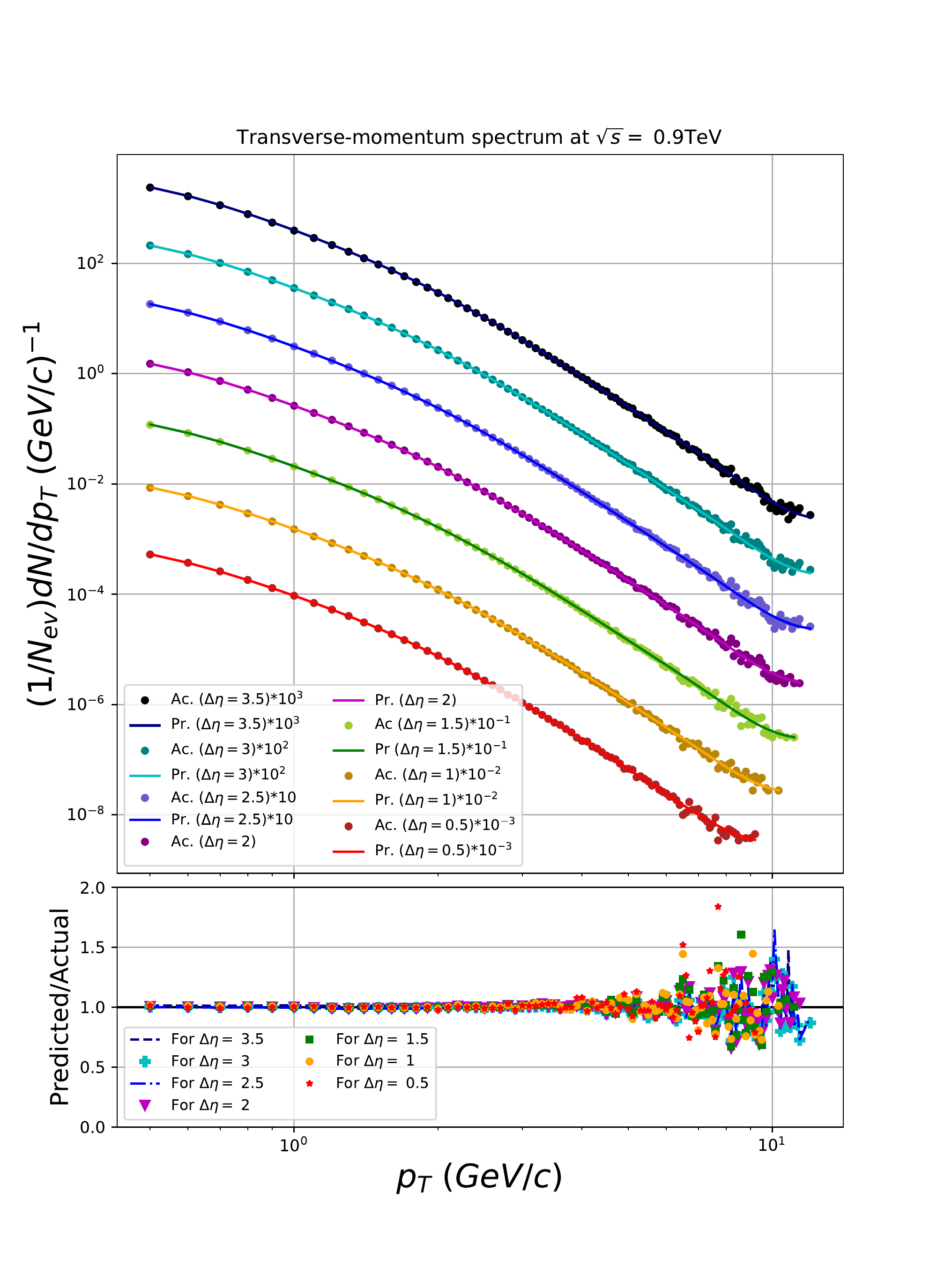}
\caption{\footnotesize for $\sqrt{s} = 0.9TeV$. } 
\end{subfigure}
  \hfill
   \begin{subfigure}{0.52\textwidth}
    \includegraphics[width=\textwidth , height=9cm]{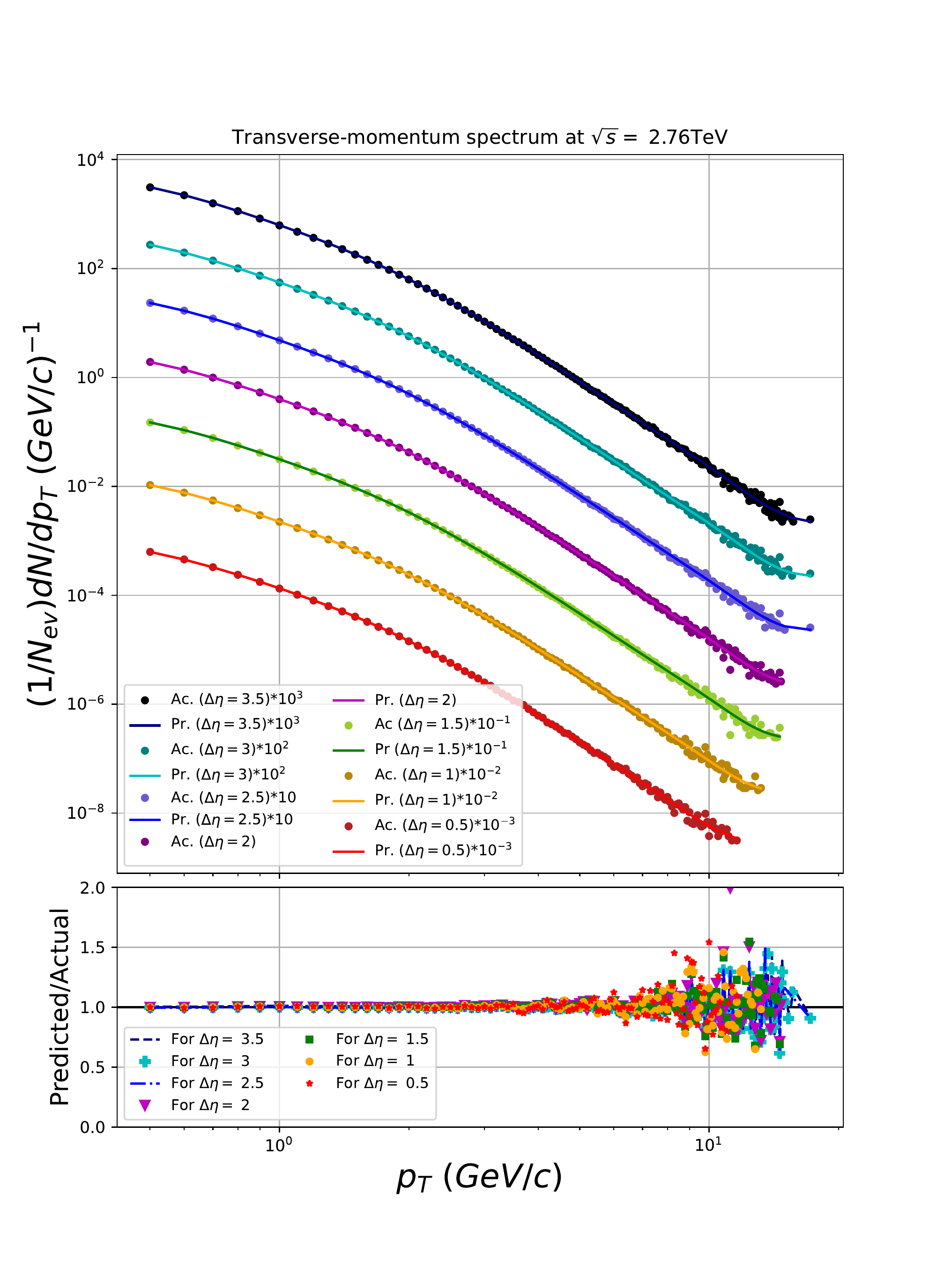}
\caption{\footnotesize  for $\sqrt{s} = 2.76TeV$. } 
\end{subfigure}

 \begin{subfigure}{0.52\textwidth}
    \includegraphics[width=\textwidth , height=9cm]{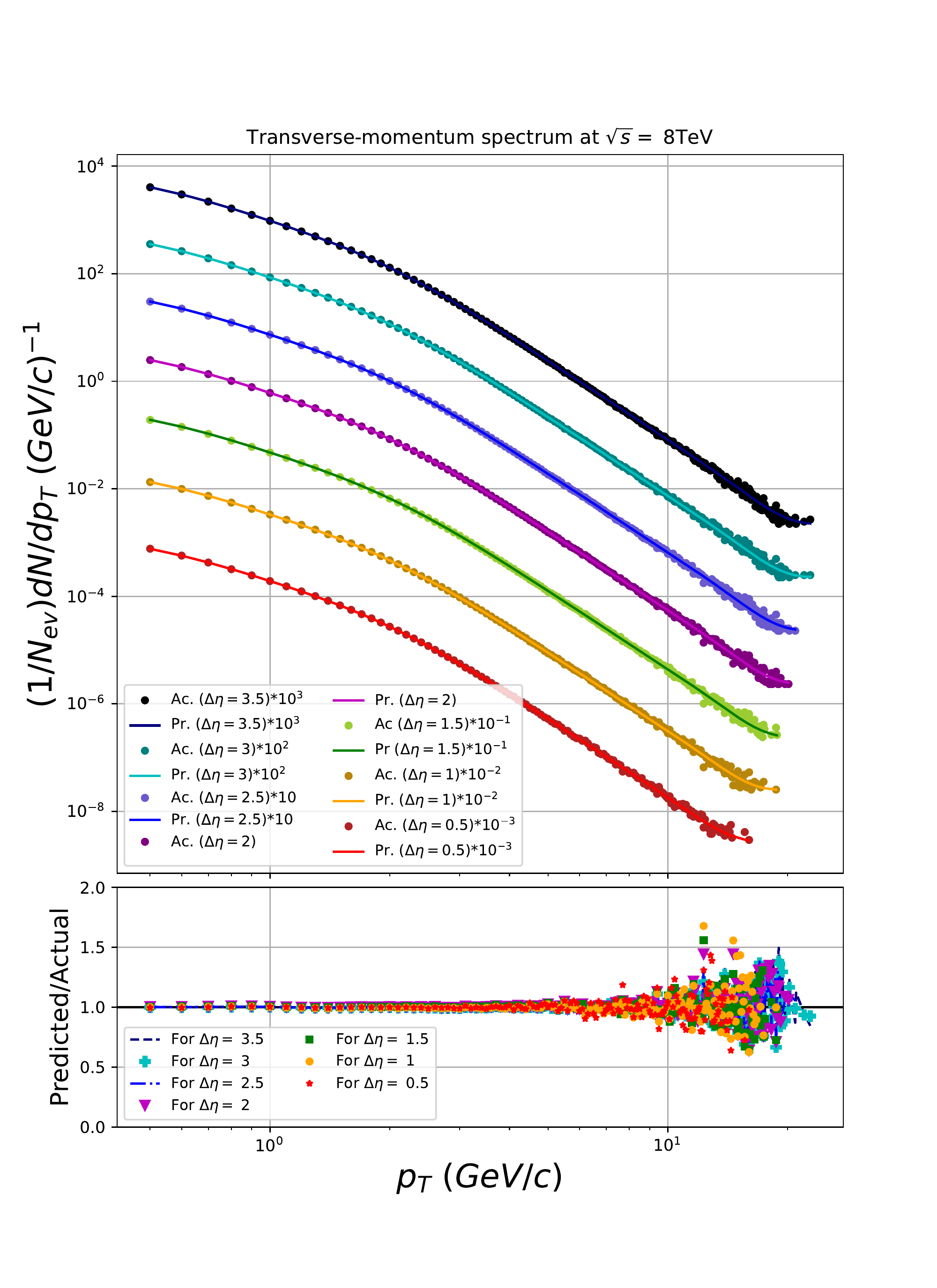}
\caption{\footnotesize for $\sqrt{s} = 8TeV$. } 
\end{subfigure}
  \hfill
   \begin{subfigure}{0.52\textwidth}
    \includegraphics[width=\textwidth , height=9cm]{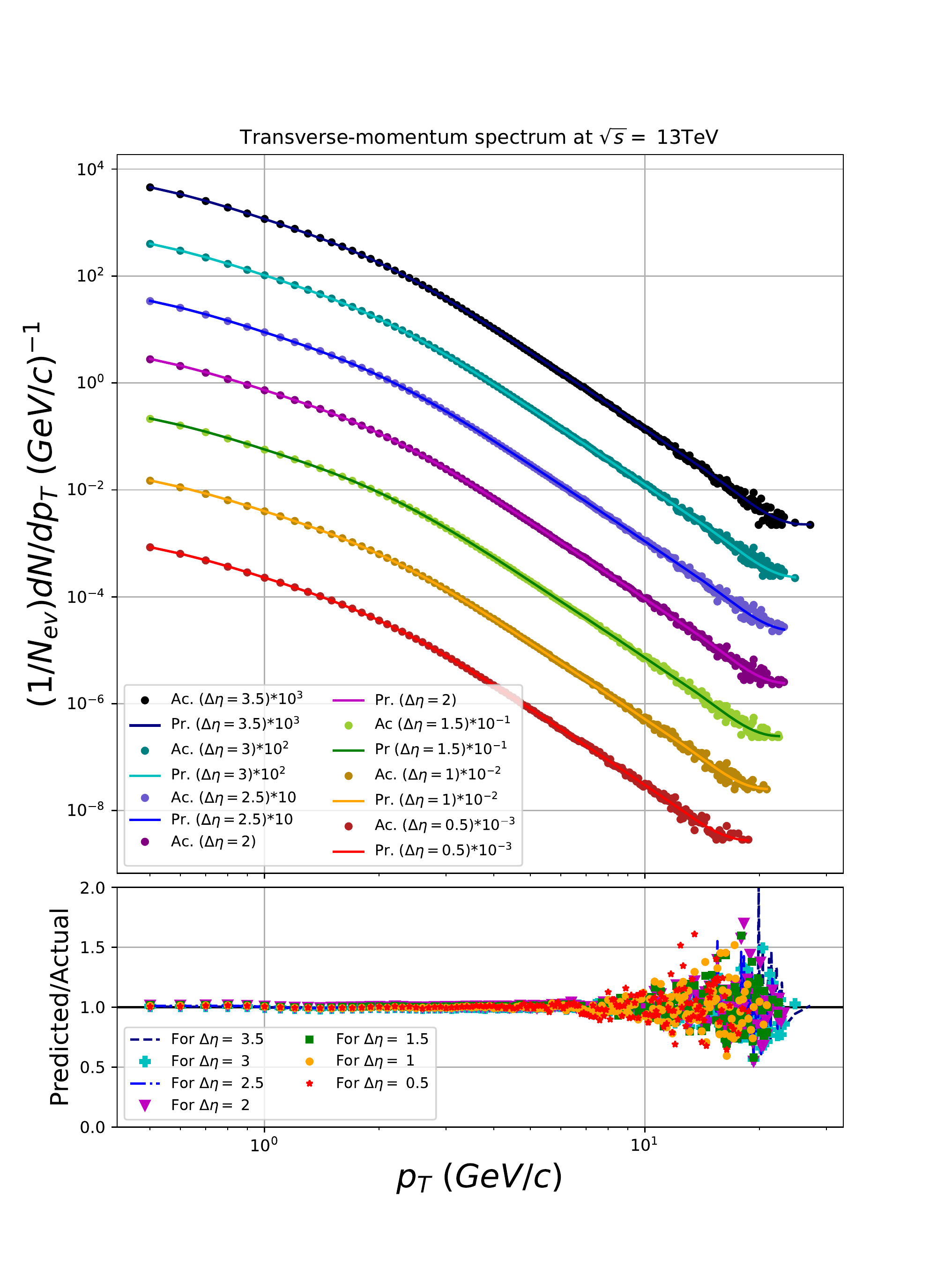}
\caption{\footnotesize for $\sqrt{s} = 13TeV$. } 
\end{subfigure}

\caption{ 
Transverse-momentum spectrum in between the Actual (Ac.) distributions generated by PYTHIA and Predicted (Pr.) by the model in case of the trained data (0.9, 2.76, 8 and 13 TeV).
} \label{pttrain}
\end{figure}

 Fig. \ref{train} and Fig. \ref{pttrain}, show 
 the comparisons of the input data with the model predictions for the CM energies used in the training, and demonstrate the quality of the model learning for the  multiplicity and transverse-momentum distributions respectively.
 For the multiplicity distributions the model correctly describes the distributions for all
 CM energies and pseudorapidity intervals. Expected
 fluctuations are seen at the high end of the 
 multiplicity distributions due to limited
  event statistics in the samples. Similarly the 
 transverse-momentum distributions are described 
 with excellent quality, in all demonstrating that 
 the DNN model used has the required flexibility.
 
The interesting part is now to check how accurate  we can "predict" distributions for different CM energies, i.e. which are not 
included in the training sets. This is checked
both for a CM energy value within the range of the training sets (10 TeV),  and for
 energy values  outside but close
 to the training 
range, and values far away from the present 
range of operation of the LHC.
As mentioned before this would be of interest for 
predictions for either possible new intermediate energy runs
of the LHC, for runs with a possible CM energy for an 
upgraded LHC, or for new future high energy colliders.
We do have to assume here that no new as yet unknown physics would set-on at these higher energies, which will significantly impact on these general inclusive variables.

\begin{figure}[!tbp]
 \begin{subfigure}{0.52\textwidth}
    \includegraphics[width=\textwidth , height=6cm]{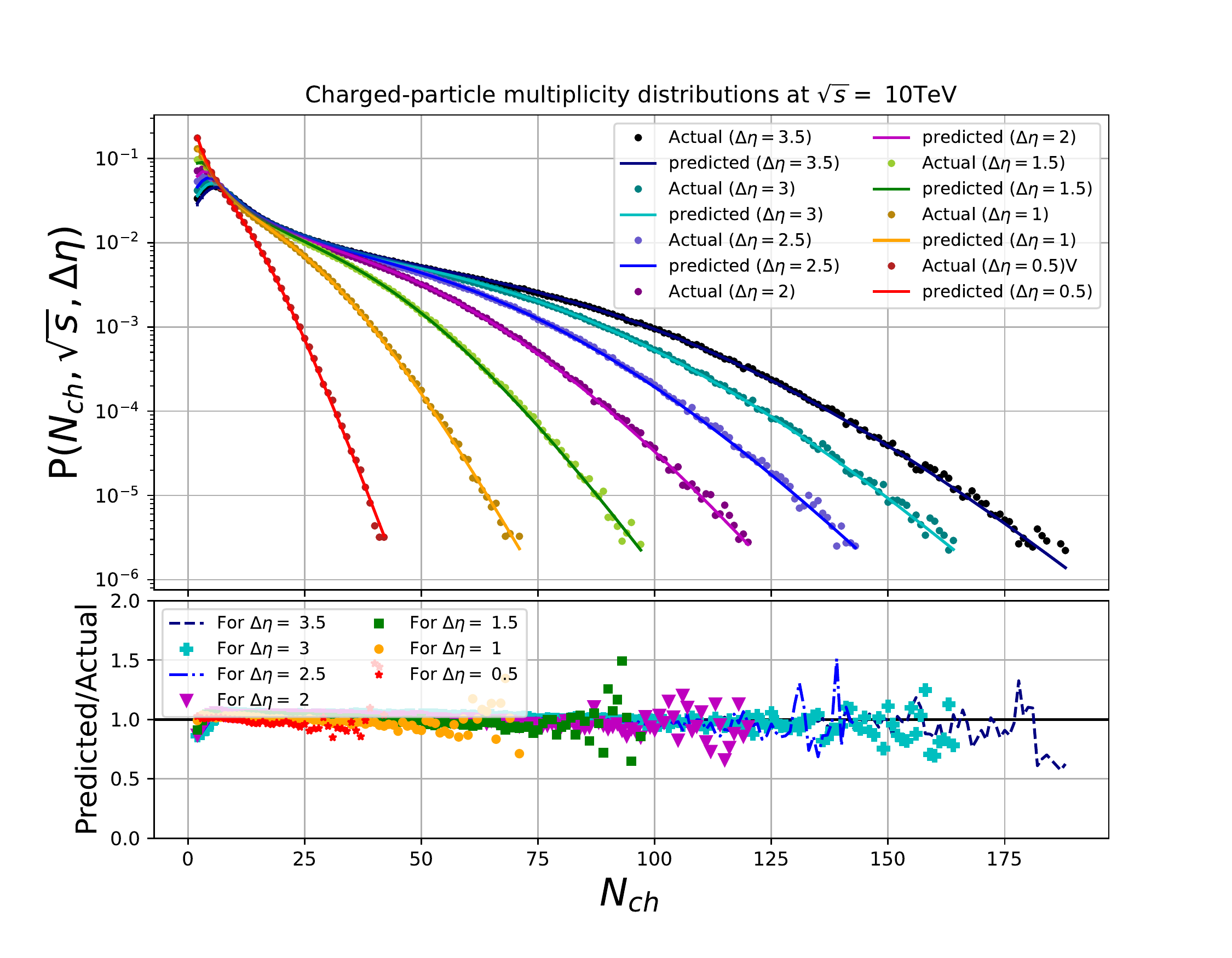}
\caption{\footnotesize for $\sqrt{s} = 10TeV$. } 
\end{subfigure}
  \hfill
   \begin{subfigure}{0.52\textwidth}
    \includegraphics[width=\textwidth , height=6cm]{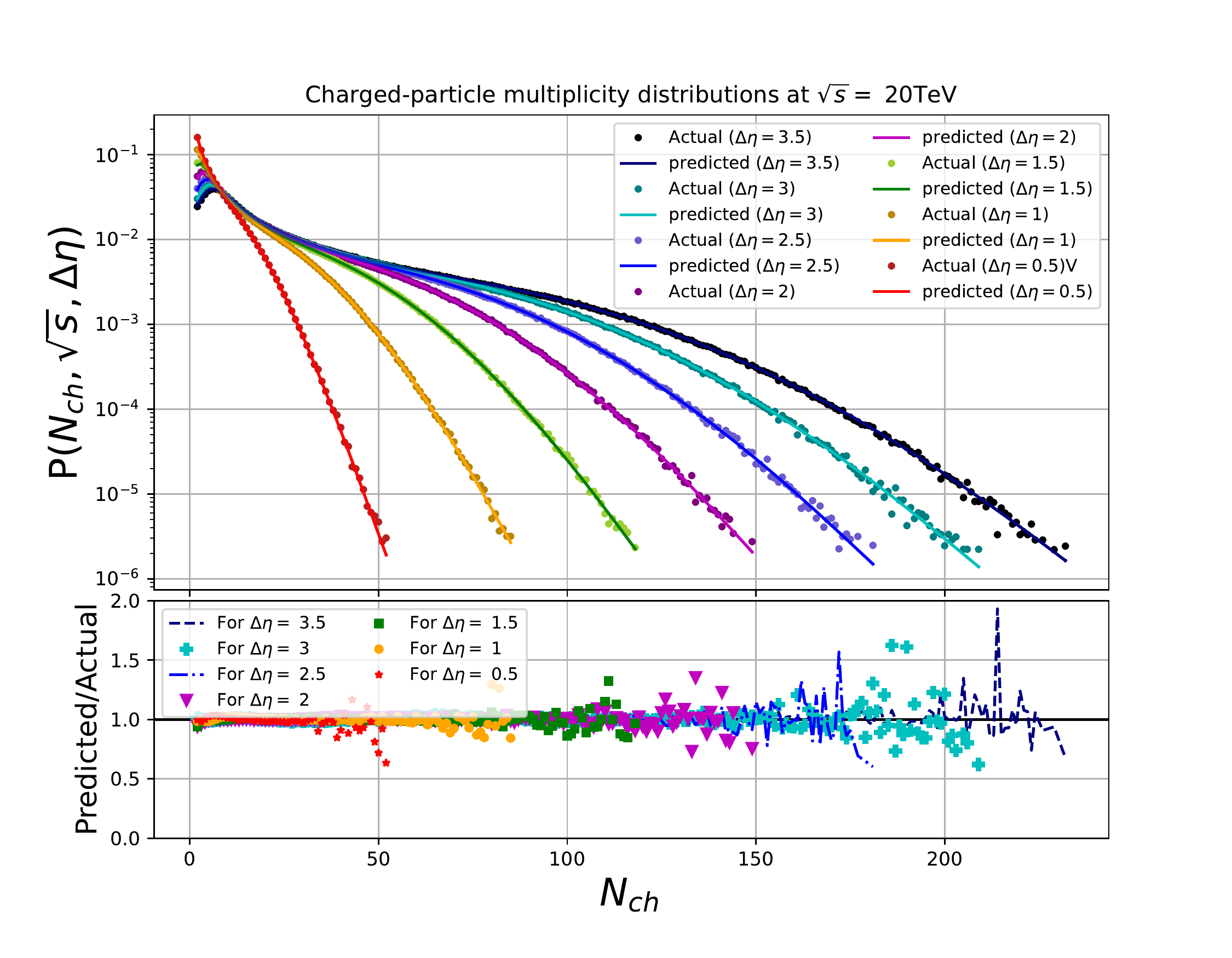}
\caption{\footnotesize  for $\sqrt{s} = 20TeV$. } 
\end{subfigure}

 \begin{subfigure}{0.52\textwidth}
    \includegraphics[width=\textwidth , height=6cm]{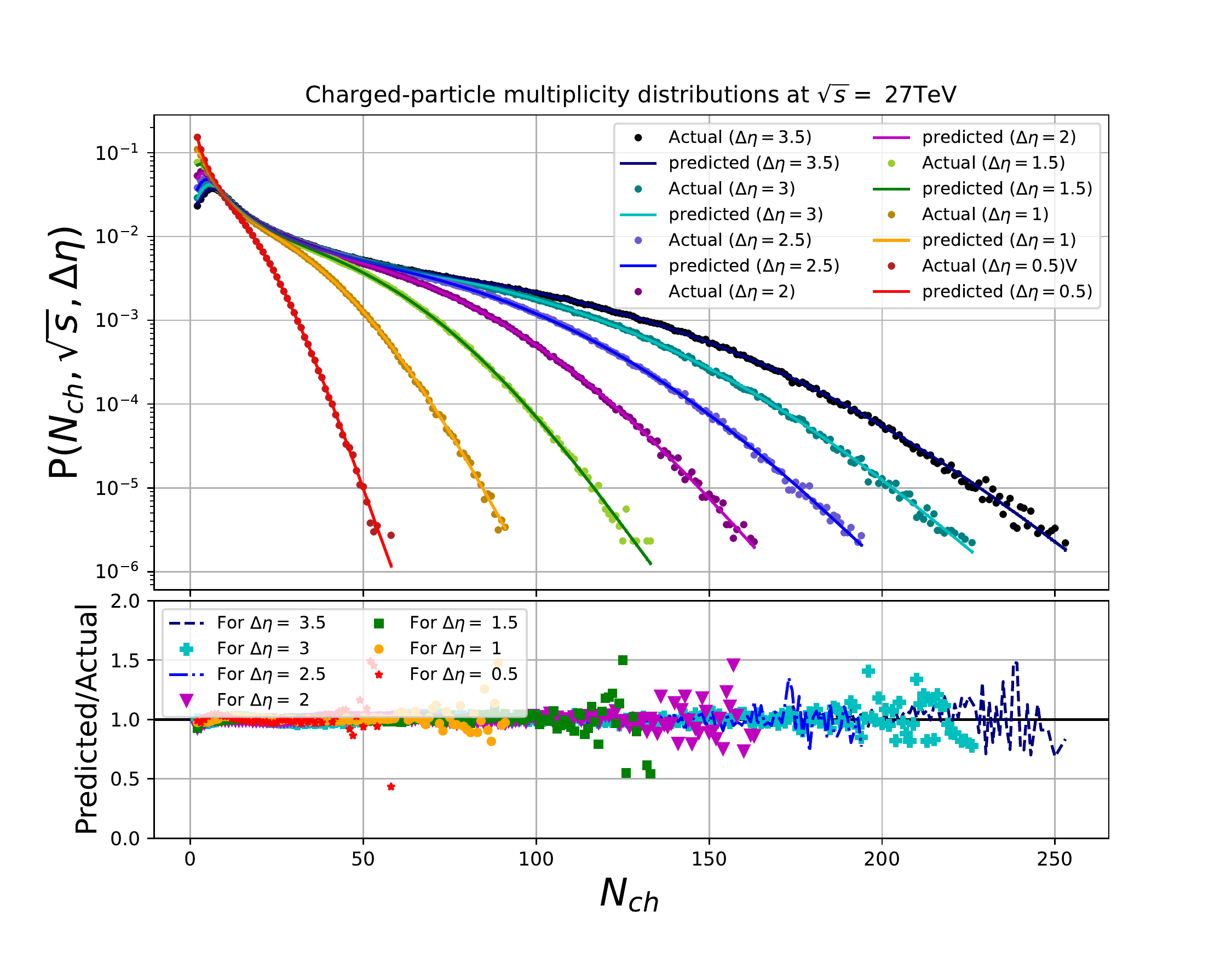}
\caption{\footnotesize for $\sqrt{s} = 27TeV$. } 
\end{subfigure}
  \hfill
   \begin{subfigure}{0.52\textwidth}
    \includegraphics[width=\textwidth , height=6cm]{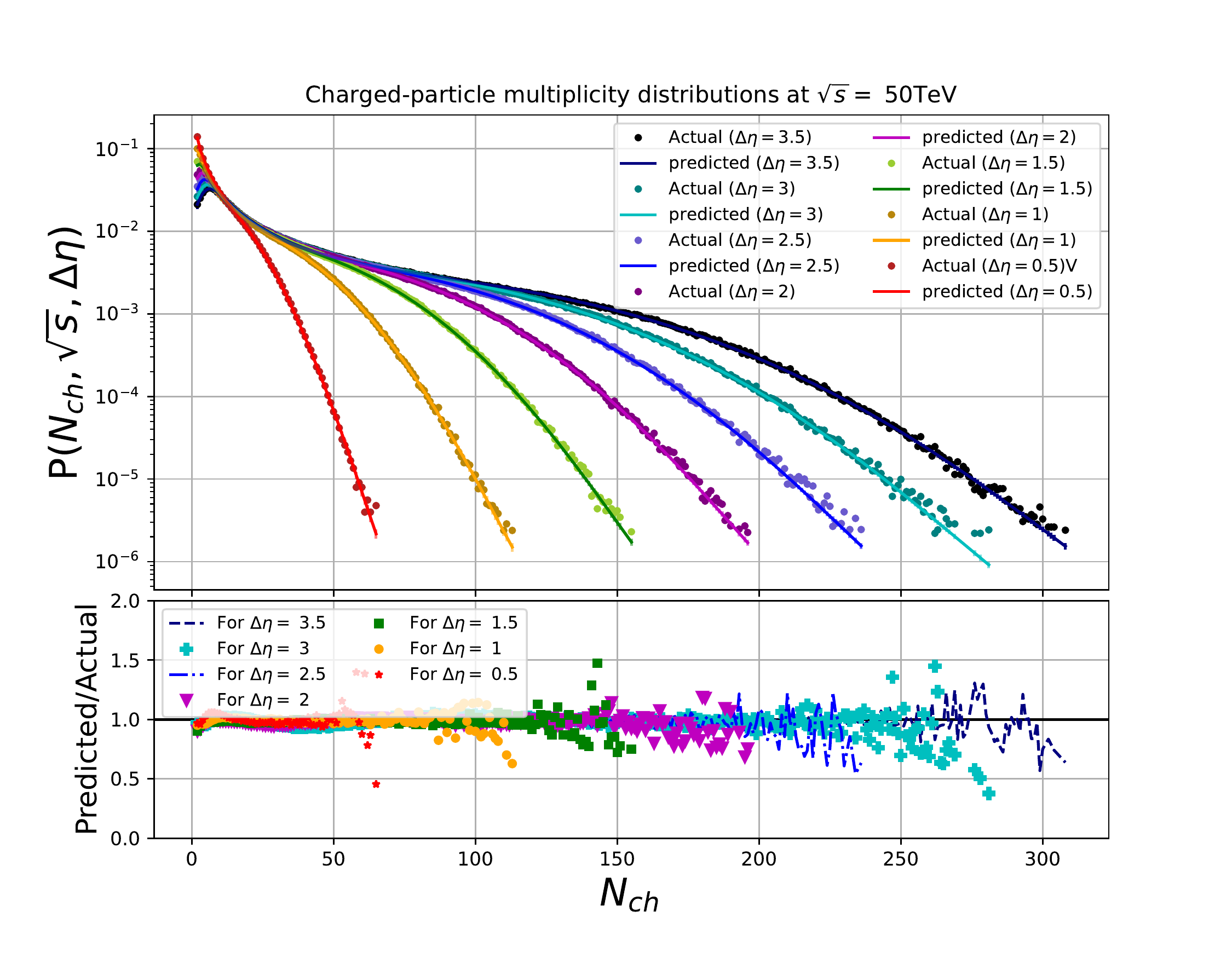}
\caption{\footnotesize for $\sqrt{s} = 50TeV$. } 
\end{subfigure}

\begin{subfigure}{0.52\textwidth}
    \includegraphics[width=\textwidth , height=6cm]{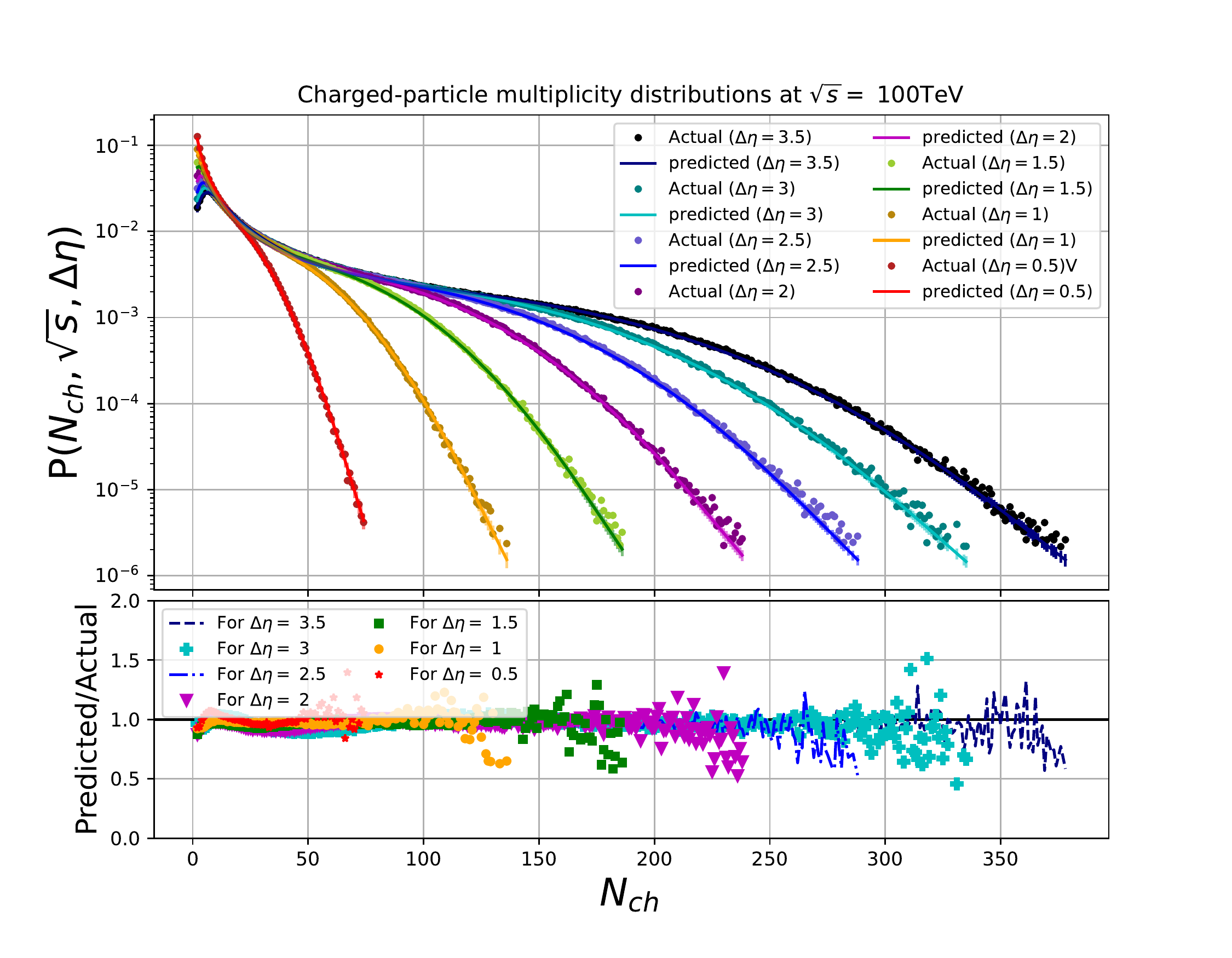}
\caption{\footnotesize for $\sqrt{s} = 100TeV$. } 
\end{subfigure}
  \hfill
   \begin{subfigure}{0.52\textwidth}
    \includegraphics[width=\textwidth , height=6cm]{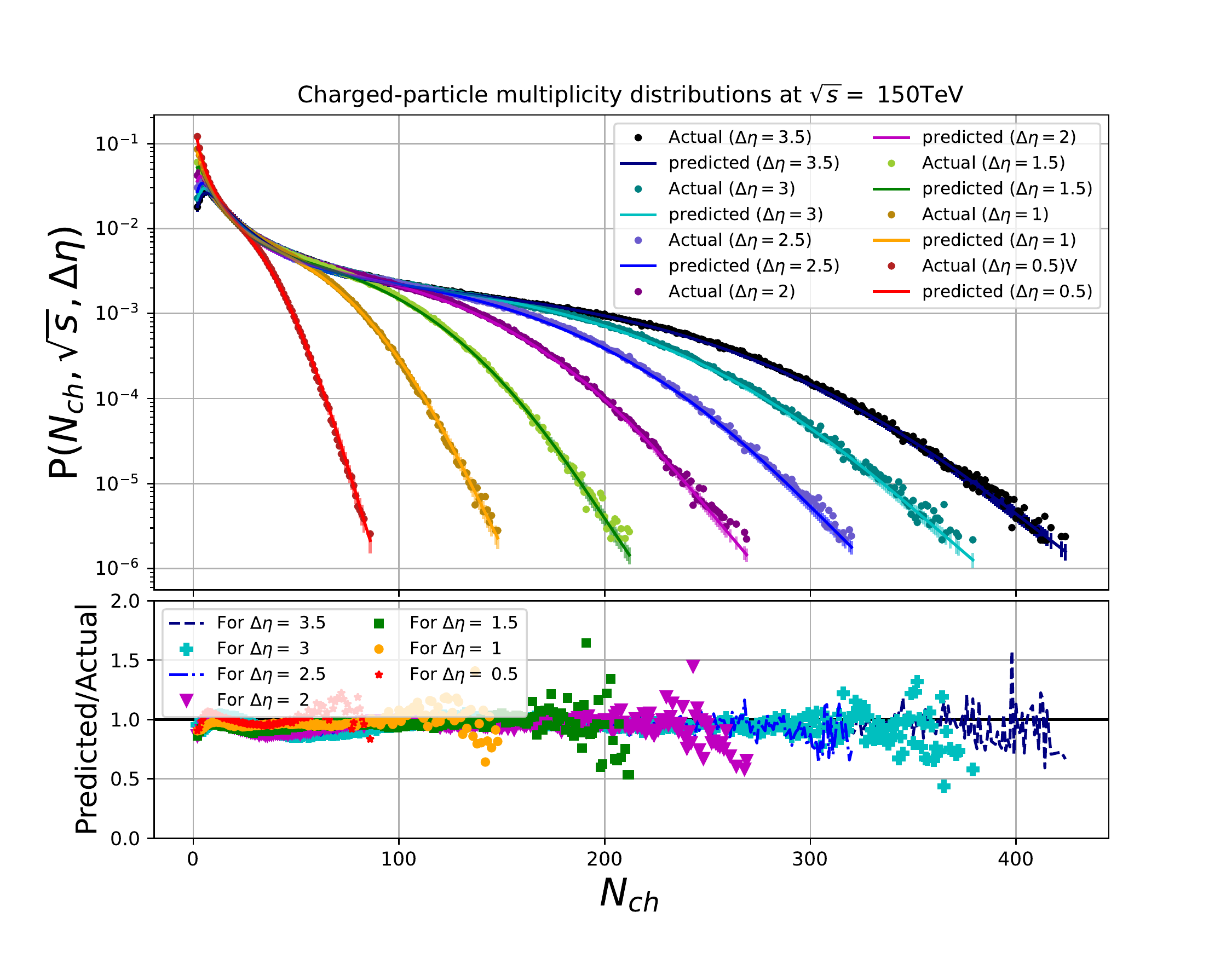}
\caption{\footnotesize for $\sqrt{s} = 150TeV$. } 
\end{subfigure}

\caption{ 
The DNN results in comparison with multiplicity distribution generated by PYTHIA for the untrained runs.
} \label{untrain}
\end{figure}

\begin{figure}[!tbp]
 \begin{subfigure}{0.52\textwidth}
    \includegraphics[width=\textwidth , height=9cm]{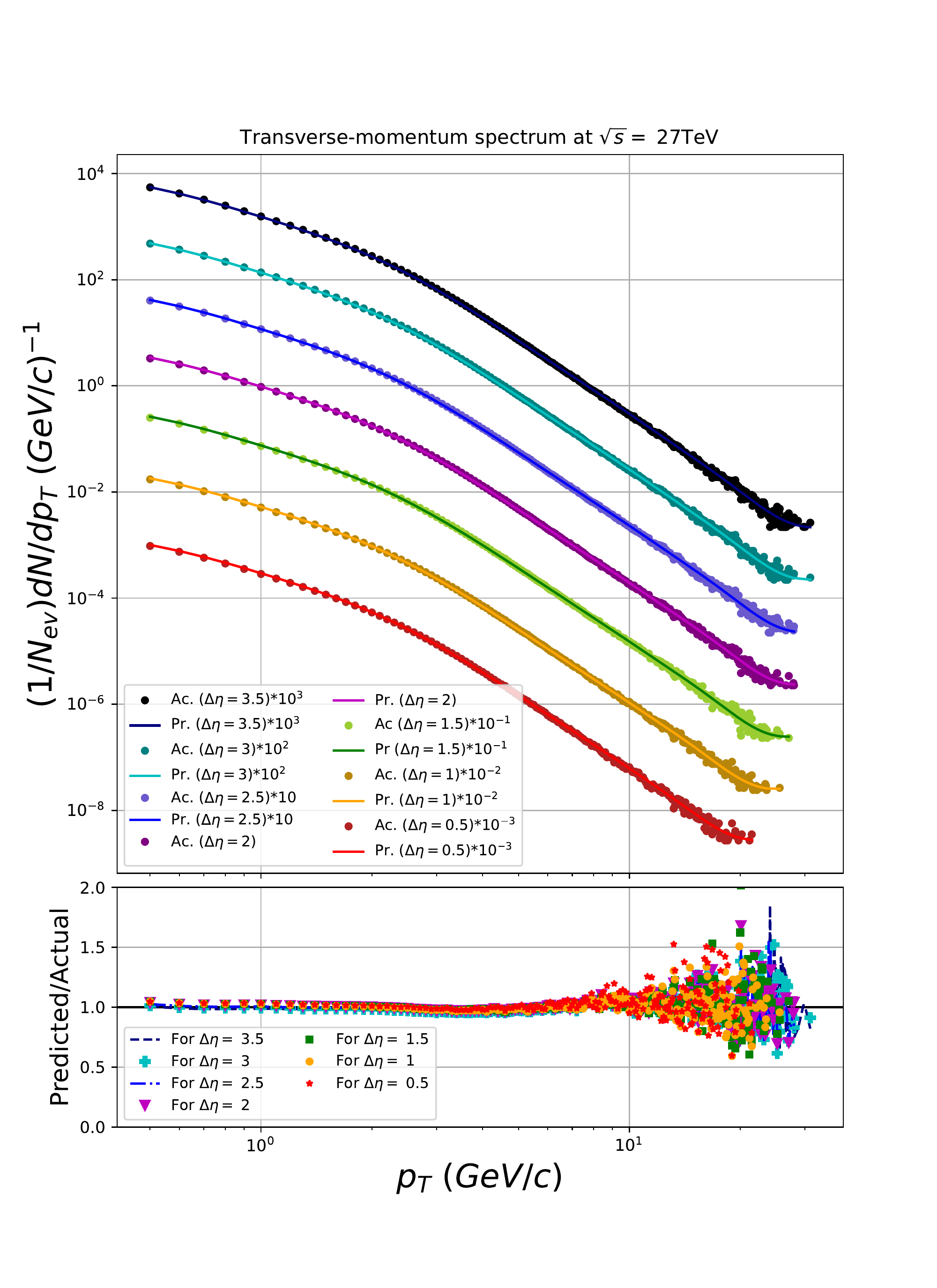}
\caption{\footnotesize for $\sqrt{s} = 27TeV$. } 
\end{subfigure}
  \hfill
   \begin{subfigure}{0.52\textwidth}
    \includegraphics[width=\textwidth , height=9cm]{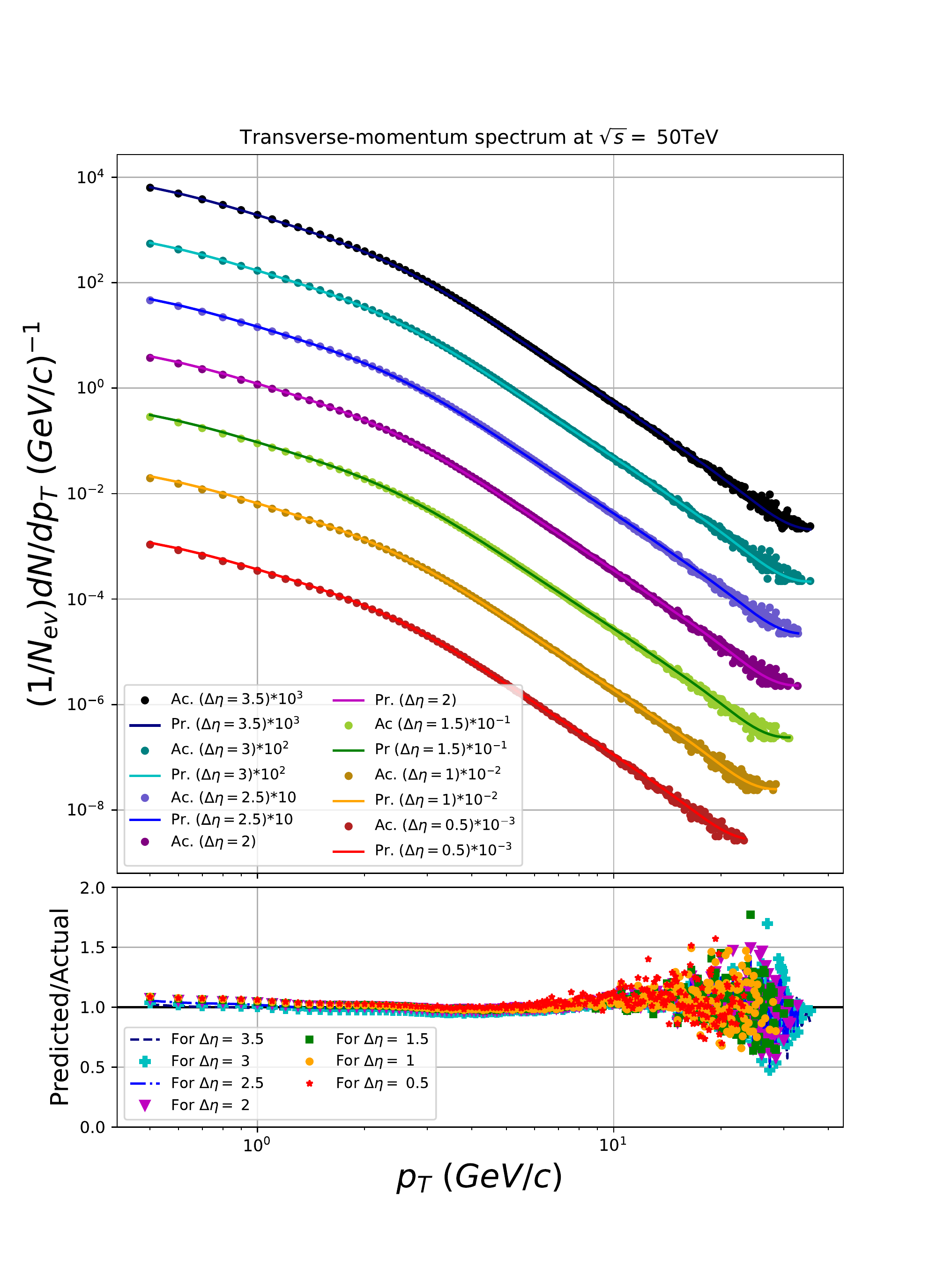}
\caption{\footnotesize  for $\sqrt{s} = 50TeV$. } 
\end{subfigure}

 \begin{subfigure}{0.52\textwidth}
    \includegraphics[width=\textwidth , height=9cm]{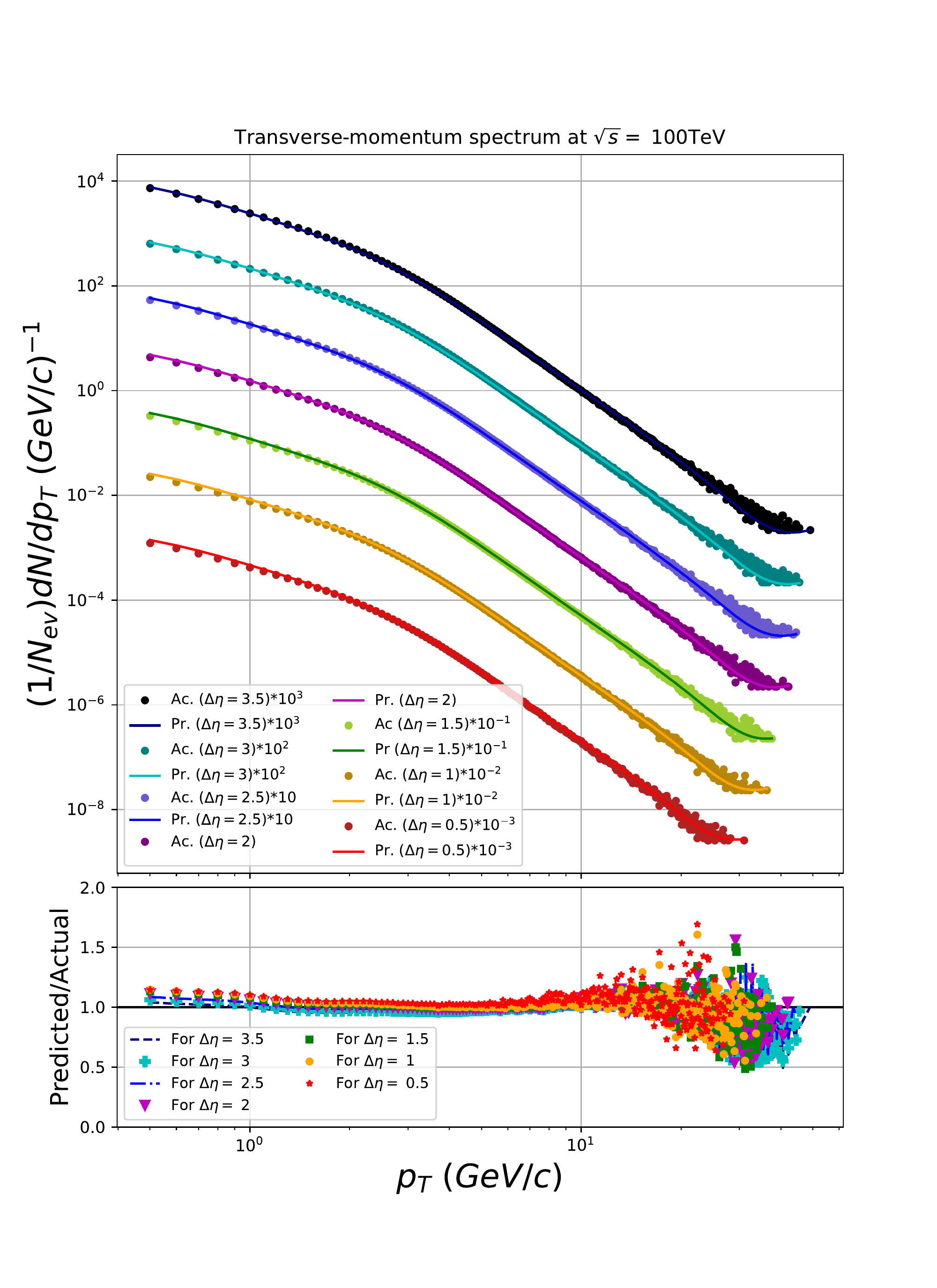}
\caption{\footnotesize for $\sqrt{s} = 100TeV$. } 
\end{subfigure}
  \hfill
   \begin{subfigure}{0.52\textwidth}
    \includegraphics[width=\textwidth , height=9cm]{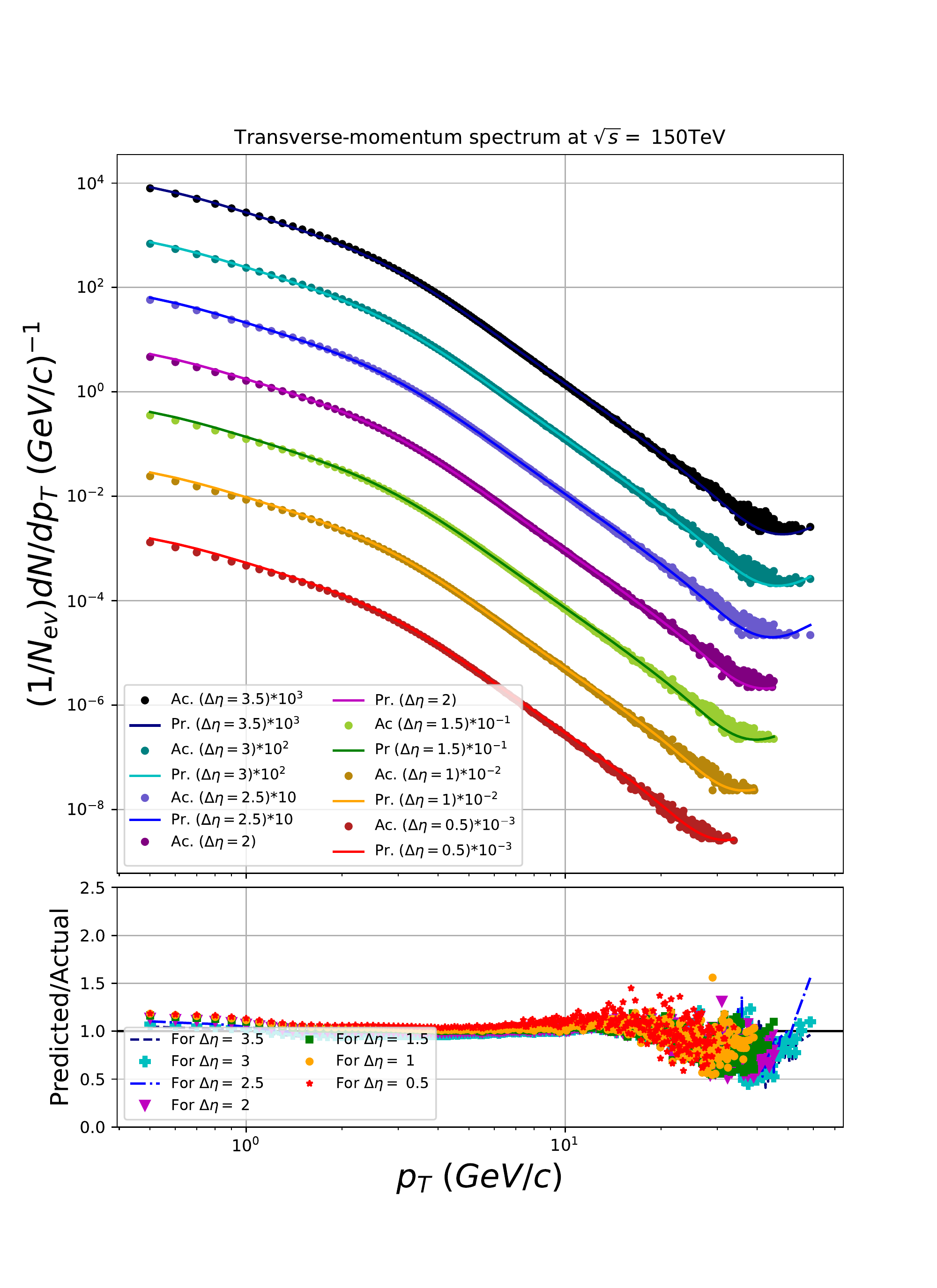}
\caption{\footnotesize for $\sqrt{s} = 150TeV$. } 
\end{subfigure}

\caption{ 
Transverse-momentum spectrum in between the Actual (Ac.) distributions generated by PYTHIA and Predicted (Pr.) by the model in case of the untrained runs (27, 50, 100 and 150TeV).} \label{ptuntrain}
\end{figure}

The result are shown in 
Figs.~\ref{untrain} and \ref{ptuntrain} and show that the model gives in general
an excellent agreement  comparing predicted  with the true PYTHIA distributions for CM energies up to 50 
TeV,  while some deviations are seen in case of highest energies tried at 100 and 150 TeV. 
For the multiplicity predictions in particular, the large 
$n_{ch}$ end the of distributions are less stable 
in that region. Similar effects are seen at the 
high $p_T$ end of the transverse momenta distributions.

In order to test the stability of our model, we have made for the multiplicity studies 50 independent tries, %
using a different splitting of the data into trained and validated sample and took the 
average of the tries as well as the envelope of the spread if the results, 
which are the curves show that on
the figures. The smallness of the 
envelope shows that the results are 
quite stable.

Furthermore, as mentioned before, we have tried a lot of different network configurations, by changing e.g. the number of layers and number of neurons per layer, different activation functions such (sigmoid, tanh) and different type of optimizers but it seems that the structure that we used in the paper shows the best predictive power.

To check the quality of the predictions we compared the normalized sum of the difference between 
predicted and observed values for the multiplicity plots. The 10 TeV prediction gives
comparable values as the ones from CM energy values used in the training, while the 
predictions for 50, 100 and 150 TeV are typically a factor 2-3 worse, but still 
of acceptable good quality.



\begin{figure}[h]
 \begin{subfigure}{0.5\textwidth}
    \includegraphics[width=\textwidth , height=7cm]{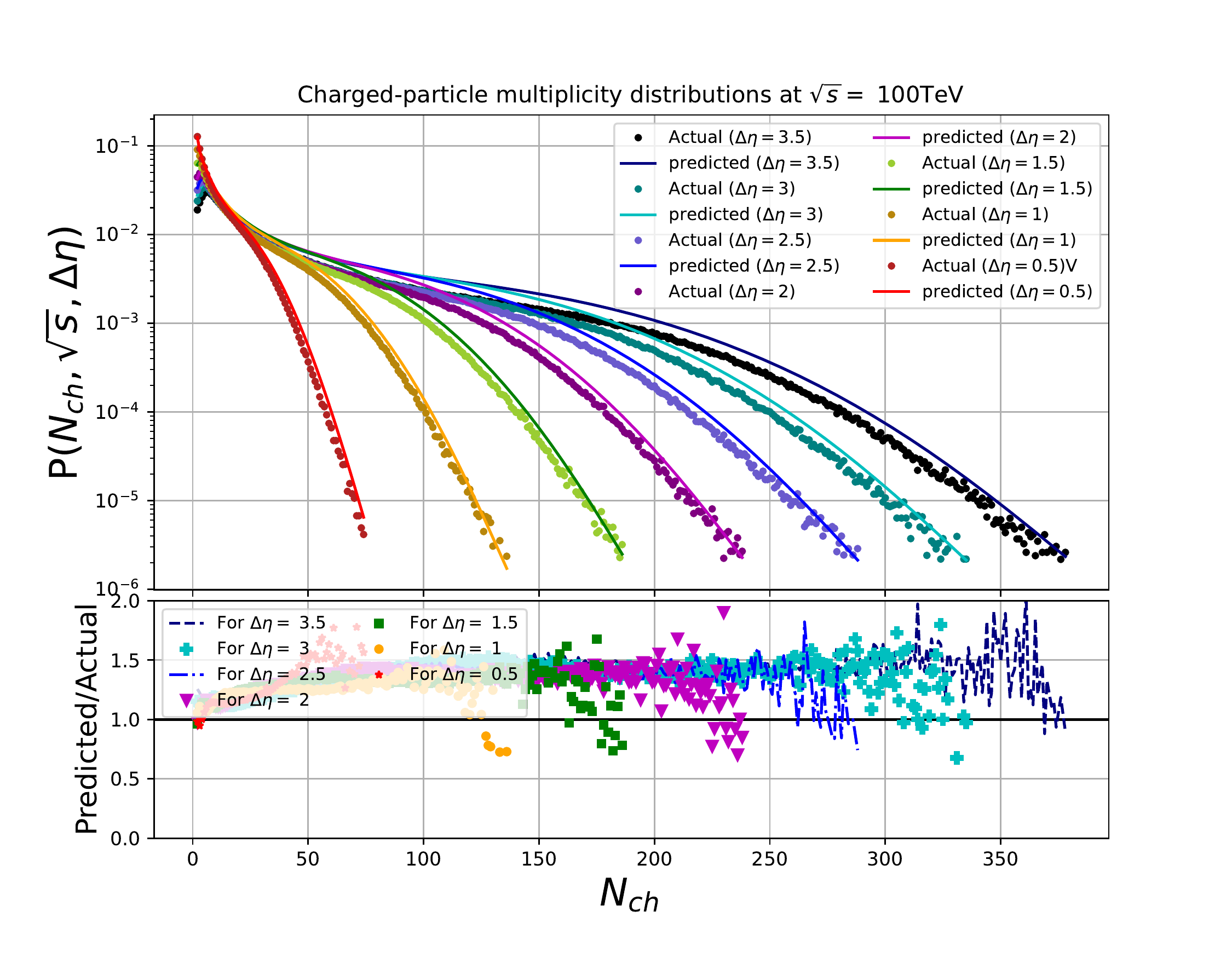}
\caption{for 7 (50m) and 13 (50m) TeV. } 
\end{subfigure}
  \hfill
   \begin{subfigure}{0.5\textwidth}
    \includegraphics[width=\textwidth , height=7cm]{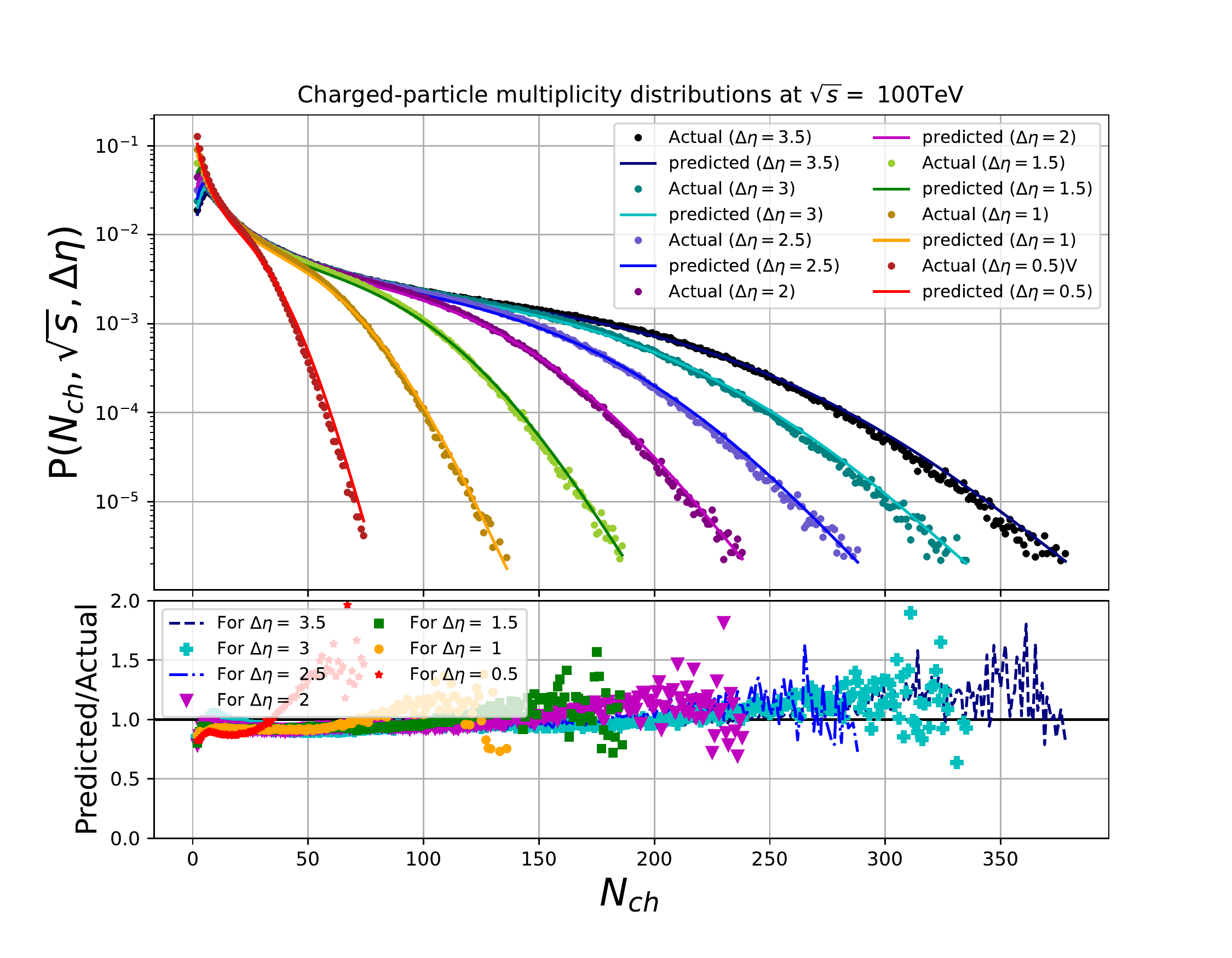}
\caption{for 2.76, 7 (50m) and 13 (50m) TeV.} 
\end{subfigure}
\caption{Test of the multiplicity model prediction at 100 TeV when training on different number of energies.}
\label{test}
\end{figure}

 A further of the stability was  made on using only two sets of energies 7 (50m) and 13 (50m) and three sets (2.76, 7 (50m) and 13 (50m) TeV) as training sets for composing the multiplicity model. We found the results are already  very stable for higher energy predictions  when using  at least three sets of separate and spread-out energy values, see Fig. \ref{test}.

The network structure of our model 
is of the form [3x20x20x1] for the structure in the different layers. We note that the output of this model can in principle directly be obtained
by multiplying the data matrices with the derived weighting matrices and adding biases for each layer, which can be represented by the following equation:

\begin{equation}
\centering
Y^{[1x1]}=f_3(f_2(f_1(X^{[3x1]}*W_1^{[20x3]}+B_1^{[1x20]})*
W_2^{[20x20]}+B_2^{[1x20]})*W_3^{[1x20]}+B_3^{[1x1]})
\end{equation}

where $Y^{[1x1]}$ is the output of our presented model, i.e. $P(N_{ch},\sqrt{s},\Delta\eta)$ in case of multiplicity and $(1/N_{ev}).dN/dp_T$ in case of $p_{T}$ modeling; $X^{[3x1]}$ is the input matrix, i.e. $N_{ch} *0.1$, $\Delta\eta$ and $\sqrt{s}$ for multiplicity and $p_{T}$, $\Delta\eta$ and $\sqrt{s}$ in case of transverse-momentum.
Here
 $f_1$,$f_2$ are the activation functions of the hidden layers which are the hyperbolic tangent functions (tanh) and $f_3$ is the activation function of the output layer, a first-order polynomial.  The matrix $W_1^{[20x3]}$ is a 20 by 3 matrix representing the weights for the first hidden layer neurons, $W_2^{[20x20]}$ is 20 by 20 matrix for the second hidden layer neurons and $W_3^{[20x1]}$ for the output layer. $B_1^{[1x20]}$ and $B_2^{[1x20]}$ are 1 by 20 matrices representing the biases for the first and second hidden layers and $B_3^{[1x1]}$ is for the output layer neuron.
 These matrices can be provided on request.

 \clearpage





\section{Conclusion}

We deploy machine learning techniques 
to build a model for the description of 
charged-particle multiplicity and transverse-momentum measurements in 
high energy $pp$ interactions.
proton-proton collisions have been generated by the event generator PYTHIA at the energies at which the LHC operated to train the  model and test its predictive 
power.
A good ML structure that shows  small loss value and high stable predictions has been reported. 

The model with the [3-20-20-1] structure, and tanh activation function in the hidden layer and a linear function for the output layer, shows an excellent agreement in comparison with the trained and untrained runs for all the seven pseudorapidity windows selected, with the coefficient of determination (see eqn. (4)) up to 0.9995 in case of multiplicity and about 0.9990 in case of $p_{T}$.

This model succeeded in providing good
predictions for the charged-particle multiplicity and transverse-momentum distributions at different center of mass energies. Hence such a procedure, when applied on real measured data at the LHC at the different energies could be used in studies for possible future CM energies, at the LHC or elsewhere, to give an
initial idea of the to be expected 
particle density in future experiments.

\section {Acknowledgment}
The Authors acknowledge the Academy of Scientific Research and Technology(ASRT), Egypt for supporting the current work under the project number 6376.

\end{document}